\documentclass[aps,prl,twocolumn,superscriptaddress,floatfix]{revtex4-2}
\usepackage{amssymb}
\usepackage{physics}
\usepackage{siunitx}
\usepackage{graphicx}
\usepackage{amsmath}
\usepackage{amsthm}
\usepackage{amsfonts}
\usepackage[T1]{fontenc}
\usepackage{array}
\usepackage{multirow}
\usepackage{color}
\usepackage{esint}
\usepackage{bm}
\usepackage{bbm}
\usepackage{hyperref}
\usepackage{babel}

\newcommand{\lettersection}[1]{\emph{#1}.---}
\newcounter{appx}
\renewcommand{\theappx}{\Alph{appx}}
\newcommand{\appxsection}[2]{\refstepcounter{appx}\label{#2}\lettersection{Appendix~\theappx: #1}}
\newcommand{\moire}{moir\'e\ }
\newcommand{\streda}{St\v{r}eda}
\hypersetup
{	colorlinks,%
	citecolor=blue,%
	linkcolor=blue,%
	urlcolor=blue%
}
\allowdisplaybreaks[4]
\begin{document}

\global\long\def\id{\mathbbm{1}}
\global\long\def\ui{\mathbbm{i}}
\global\long\def\ud{\mathrm{d}}

\title{Interaction-induced sign reversal of the orbital magnetic susceptibility in Chern insulators}

\author{Ke Huang}
\affiliation{Department of Physics, City University of Hong Kong, Kowloon, Hong Kong SAR, China}

\author{Xiao Li}
\email{xiao.li@cityu.edu.hk}
\affiliation{Department of Physics, City University of Hong Kong, Kowloon, Hong Kong SAR, China}

\date{\today}

\begin{abstract}
It has been well established that the orbital magnetization of interacting electrons can be simply evaluated by applying the single-particle formula to self-consistent Hartree-Fock (HF) bands. 
However, we show that such a procedure fails qualitatively for orbital magnetic susceptibility, especially in topological systems: in a Chern-insulating phase of twisted MoTe$_2$, the interaction-induced correction reverses the sign of the susceptibility. 
This result follows from an algebraic framework that solves the HF problem at finite magnetic field, where noncommuting canonical momenta obstruct a direct calculation: a \emph{reverse Peierls substitution} maps every magnetic-translation-invariant operator to a unique bivariate function, converting the finite-field self-consistency into a function equation that can be expanded systematically in $B$. 
At first order, this yields a linear equation for the field-induced change $\delta X$ of the Fock potential, leading to an intrinsically interaction-induced susceptibility in addition to a single-particle-like one. 
The framework reproduces the \streda\ formula for insulators, and an auxiliary-Hilbert-space construction carries it to periodic and \moire systems. 
Finite-field HF calculations in a gapped Dirac model and in the twisted MoTe$_2$ Chern insulator confirm the theory quantitatively. 
\end{abstract}

\maketitle

\lettersection{Introduction}
Orbital magnetism has emerged as a central diagnostic of correlated topological phases in \moire and crystalline two-dimensional materials, aided by advances in scanning SQUID-on-tip microscopy that provide nanoscale spatial resolution and high magnetic sensitivity under cryogenic and high-field conditions~\cite{Finkler2012,Vasyukov2013,Anahory2020,ZhouSOT2023}. In these systems, interaction-driven Chern insulators, orbital ferromagnets, and fractional Chern insulators---first identified through anomalous Hall transport and thermodynamic measurements~\cite{Sharpe2019,Serlin2020,Cai2023,Zeng2023,Park2023,Xu2023}---carry orbital magnetic moments that are now directly resolved by nanoscale magnetic imaging~\cite{Tschirhart2021,Grover2022,Redekop2024,Auerbach2025}.

The orbital magnetic susceptibility---the leading response of the magnetization to a weak field---encodes the geometry and topology of the underlying bands and, in interacting systems, the response of the self-consistent mean field itself. 
Moreover, local imaging of magnetization at very weak magnetic fields~\cite{ZhouPMF2023} opens up the possibility of detecting magnetic susceptibility due to orbital contributions. 
A predictive, gauge-invariant theory of the susceptibility for interacting electrons is therefore essential for interpreting these measurements. 
Indeed, we show below that interactions can reverse the sign of the orbital susceptibility of a Chern insulator in twisted MoTe$_2$ (tMoTe$_2$)---an effect invisible to the single-particle formula even when it is evaluated on the self-consistent Hartree-Fock bands, the very procedure that succeeds for the orbital magnetization~\cite{ShiVignaleXiaoNiu2007,Vignale2026}.

For noninteracting electrons the orbital susceptibility is well understood, from the effective-Hamiltonian and semiclassical treatments of Bloch electrons in a magnetic field~\cite{Blount1962,Roth1966} to the modern theory of orbital magnetization and its geometric corrections~\cite{XiaoShiNiu2005,Thonhauser2005,ShiVignaleXiaoNiu2007,XiaoChangNiu2010,Gao2015}. 
Interactions are commonly treated within the Hartree-Fock approximation, which at zero field reduces to a self-consistent function of the Bloch momentum and---as noted above---leaves the orbital magnetization in its single-particle form once evaluated on the self-consistent quasiparticle bands~\cite{ShiVignaleXiaoNiu2007}, a statement recently shown to hold exactly in current-density-functional theory as well~\cite{Vignale2026}. 
At finite field this Bloch-momentum description is lost: the canonical momenta no longer commute, the system discretizes into Landau levels, and a direct finite-field HF calculation is numerically demanding. 
Recent work by Kang, Wang, and Vafek~\cite{Kang2025} made important progress by deriving an HF expression for the orbital magnetization and presenting a susceptibility formula written entirely in terms of the zero-field solution, numerically tested both against finite-field calculations in a translation-invariant Rashba model with a contact interaction; the detailed derivation of the susceptibility, however, was deferred to a future publication, and the numerical validations in more generic systems were left open. 
This is part of a broader recent effort on orbital magnetism of interacting electrons, including exact response-function formulations of the magnetization~\cite{ChenSong2026}, Luttinger-Ward-based field expansions for correlated multiband systems~\cite{Ye2026}, finite-field mean-field studies of quantum Hall ferromagnets~\cite{Huang2026}, and Hartree-Fock calculations of the magnetization of correlated states in twisted transition-metal dichalcogenides~\cite{Liu2026}; none of these, however, addresses the magnetic susceptibility itself.

In this Letter we address the susceptibility directly through an algebraic reformulation of the finite-field HF problem.
The key object is the reverse Peierls substitution $\mathcal{R}$, which assigns to any operator commuting with the guiding-center momentum a unique bivariate function; because $\mathcal{R}$ is linear, the HF self-consistency becomes a function equation valid at both zero and finite field, and the noncommutativity enters only through a controlled $\order{B}$ expansion of operator products. Solving this equation to first order gives a linear equation for the change $\delta X$ of the Fock potential that is efficient to solve and directly reusable. 
From it we obtain both $\pdv*{n}{B}$---which reduces to the \streda\ formula for insulators---and the susceptibility $\chi^{\text{HF}}=\tilde\chi_{\delta X}+\chi^{\text{single}}_{H_0^{\text{MF}}}$, cleanly separating an interaction-induced piece from a single-particle-like one. A mapping to an extended Hilbert space renders periodic and \moire systems formally translation invariant, so the same equations apply. We validate the theory against self-consistent finite-field HF in two models, extracting $\chi$ two independent ways, and show in a tMoTe$_2$ Chern insulator that the interaction correction can flip the sign of the susceptibility---demonstrating that the single-particle formula alone is qualitatively inadequate for interacting topological bands.

\lettersection{Hartree-Fock self-consistent equation}
An interacting Hamiltonian of a 2D system with translation symmetry has the following generic form at zero magnetic field,
\begin{align}
	H=h(\hat{\vb k})+&\sum_{\alpha,\beta}\frac{1}2\int\dd[2]{r}\dd[2]{r'}V(\vb r-\vb r')\nonumber\\
	& \times\psi^\dag_{\alpha}(\vb r)\psi^\dag_{\beta}(\vb r')\psi_{\beta}(\vb r')\psi_{\alpha}(\vb r),
\end{align}
where $V(r)$ is the interaction potential, and $\alpha,\beta$ denote the other degrees of freedom, such as sublattices, valleys, or even nonlocal degrees of freedom, which will be elaborated on below. The single-particle Hamiltonian $h(\hat{\vb k})$ is a matrix-valued bivariate function, and its entries describe the tunneling between different degrees of freedom.

In the presence of a finite and uniform perpendicular magnetic field, the above Hamiltonian is modified by substituting the momentum operator $\hat{\vb k}$ with the canonical momentum operator $\hat{\vb*\pi}$, whose components no longer commute, $[\hat\pi_x,\hat\pi_y]=-iB$. 
A bivariate function of $\hat{\vb*\pi}$ is therefore defined only once an operator ordering is specified. Setting $\hbar=e=1$, we adopt throughout the Weyl (symmetric) ordering,
\begin{align}\label{Eq:Weyl}
	h(\hat{\vb*\pi})=\int\dd[2]{s} h^{\mathcal F}(\vb s)e^{i\vb s\vdot \hat{\vb*\pi}}
\end{align}
with $h^{\mathcal F}(\vb s)$ denoting the Fourier transform of $h(\vb k)$. 
This is the standard choice for lattice models: $e^{i\vb s\vdot\hat{\vb*\pi}}$ translates by $\vb s$ while accumulating the Peierls phase along the straight-line path joining the two endpoints, so whenever $h^{\mathcal F}(\vb s)$ is supported on the Bravais lattice vectors, Eq.~\eqref{Eq:Weyl} is precisely the conventional Peierls substitution applied to the hopping amplitudes. Weyl ordering further maps Hermitian $h(\vb k)$ to Hermitian $h(\hat{\vb*\pi})$. All statements below---in particular the uniqueness of the reverse Peierls substitution introduced next---are understood with respect to this convention.

In the Hartree-Fock (HF) approximation, the mean-field Hamiltonian is given by (see Appendix~\ref{App:selfcons})
\begin{align}
	\hat{H}^{\text{MF}}&=h(\hat{\vb*\pi})+\hat X,\label{Eq:HMF}\\
	\hat X&\equiv V_{\text{MF}}[\hat N]:=-\int\frac{\dd[2]{q}}{(2\pi)^2}\tilde V(q)e^{i\vb q\vdot \hat{\vb r}}\hat N e^{-i\vb q\vdot \hat{\vb r}},\label{Eq:Fock}
\end{align}
where $\tilde V(q)$ is the interaction potential in the momentum space, $\hat X$ is the HF correction, and $\hat N=f_{\text{FD}}(\hat{H}^{\text{MF}})$ is the reduced density matrix of the HF ground state with $f_{\text{FD}}(\varepsilon)$ denoting the Fermi-Dirac distribution. 
Here, we assume the (magnetic) translation symmetry, so the Hartree term becomes a constant potential and has been discarded in Eq.~\eqref{Eq:HMF}. The mean-field potential $V_{\text{MF}}[\hat N]$ of Eq.~\eqref{Eq:Fock} therefore contains only the Fock term, and is manifestly linear in the density matrix.

At zero field, translation symmetry guarantees that $\hat X$ is a function of the momentum operator. 
At finite field, $\hat X$ instead respects magnetic translation symmetry, $[\hat X,\hat{\vb Q}]=0$, where $\hat{\vb Q}=\hat{\vb*\pi}+B\,\hat{\vb r}\cp\vb z$ is the guiding-center momentum that generates magnetic translations.
The key observation is that this weaker symmetry is equally constraining: as we prove in Appendix~\ref{App:peierls}, any operator $\hat A$ commuting with $\hat{\vb Q}$ can be written as $\hat A=A(\hat{\vb*\pi})$ with a unique bivariate function $A(\vb k)$.
Recovering $A(\vb k)$ from $\hat A$ undoes the substitution of Eq.~\eqref{Eq:Weyl}; we call this linear map the \emph{reverse Peierls substitution}, $\mathcal{R}(\hat A):=A(\vb k)$.
Because $\hat H^{\text{MF}}$ and $\hat N$ both commute with $\hat{\vb Q}$, they are functions of $\hat{\vb*\pi}$, and the HF correction becomes
\begin{align}
	V_{\text{MF}}[N(\hat{\vb*\pi})]&=-\int\frac{\dd[2]{q}}{(2\pi)^2}\tilde V(q)e^{i\vb q\vdot \hat{\vb r}}N(\hat{\vb*\pi})e^{-i\vb q\vdot \hat{\vb r}}\nonumber\\
	&=-\int\frac{\dd[2]{q}}{(2\pi)^2}\tilde V(q)N(\hat{\vb*\pi}-\vb q).\label{Eq:Fockpi}
\end{align}
Applying $\mathcal{R}$ to Eq.~\eqref{Eq:Fockpi} and to $\hat N=f_{\text{FD}}(\hat{H}^{\text{MF}})$ gives the self-consistent equations, valid at both zero and finite fields,
\begin{align}
	X(\vb k)&=-\int\frac{\dd[2]{q}}{(2\pi)^2}\tilde V(q)N(\vb k-\vb q)\equiv V_{\text{MF}}[N(\vb k)]\label{Eq:SC0},\\
	N(\vb k)&=\mathcal{R}\{f_{\text{FD}}[h(\hat{\vb*\pi})+X(\hat{\vb*\pi})]\}\label{Eq:SC2},
\end{align}
where $X(\vb k)$ and $N(\vb k)$ are to be solved. 
Note that Eq.~\eqref{Eq:SC0} is now a function equation rather than an operator equation.

\lettersection{Magnetic correction to the density matrix}
To calculate magnetic susceptibility, we need to understand how the self-consistent solution changes with magnetic fields.
Let $X_B(\vb k),N_B(\vb k)$ be the self-consistent solution at magnetic field $B$, with mean-field Hamiltonian $H_B^{\text{MF}}(\vb k)$. At zero field the momentum components commute, so Eq.~\eqref{Eq:SC2} reduces to $N_0(\vb k)=f_{\text{FD}}[H_0^{\text{MF}}(\vb k)]$. At finite fields, $(X_0,N_0)$ continues to satisfy Eq.~\eqref{Eq:SC0} but fails to satisfy Eq.~\eqref{Eq:SC2}, because $\mathcal{R}[f(\hat A)]\neq f[\mathcal{R}(\hat A)]$ for nonlinear $f$ and generic $\hat A$.
The subtlety caused by the noncommutativity at finite fields can be demonstrated by considering the product of two generic functions of $\hat{\vb*\pi}$,
\begin{align}
	&\quad F(\hat{\vb*\pi})G(\hat{\vb*\pi})\nonumber\\
	&=\int\dd[2]{s}\dd[2]{s'} F^{\mathcal F}(\vb s)G^{\mathcal F}(\vb s')e^{i\vb s\vdot \hat{\vb*\pi}}e^{i\vb s'\vdot \hat{\vb*\pi}}\nonumber\\
	&=\int\dd[2]{s}e^{i\vb {s}\vdot \hat{\vb*\pi}}\int\dd[2]{s'} F^{\mathcal F}(\vb s-\vb s')G^{\mathcal F}(\vb s')e^{iB\vb z\vdot (\vb{s}\cp\vb s')/2}.
\end{align}
By expanding this equation to $\order{B^2}$, we obtain 
\begin{align}\label{Eq:algebra}
	&\quad \mathcal{R}[F(\hat{\vb*\pi})G(\hat{\vb*\pi})]\nonumber\\
	&=F(\vb k)G(\vb k)-\dfrac{iB}{2}\vb z\vdot[\partial F(\vb k)\cp \partial G(\vb k)]+\order{B^2},
\end{align}
which yields an additional term at finite fields. Equation~\eqref{Eq:algebra} is the $\order{B}$ truncation of the Moyal star product~\cite{Groenewold1946,Moyal1949} associated with the Weyl ordering of Eq.~\eqref{Eq:Weyl}; for noninteracting Bloch electrons, expansions of this type underlie the classic semiclassical theory of the susceptibility~\cite{Blount1962,Roth1966}, and a related noncommutative expansion has recently been applied to the orbital magnetization of correlated electrons~\cite{Ye2026}, though not to the susceptibility considered here.


Using Eq.~\eqref{Eq:algebra} recursively, the linear-in-$B$ coefficient of the density matrix caused by the noncommutativity is 
\begin{align}\label{Eq:dNpi}
	\delta N_{\pi}(\vb k):=\lim_{B\to 0}\frac{\mathcal{R}[f_{\text{FD}}(H_0^{\text{MF}}(\hat{\vb*\pi}))]-f_{\text{FD}}[H_0^{\text{MF}}({\vb k})]}B
\end{align}
can be explicitly calculated, as detailed in Appendix~\ref{App:density}.
A nonvanishing $\delta N_{\pi}$ in turn shifts the Fock potential itself, $X_B(\vb k)=X_0(\vb k)+B\delta X(\vb k)+\order{B^2}$. Consequently, the total change of the density matrix is
\begin{equation*}
N_B(\vb k)-N_0(\vb k)=B\delta N_{\pi}(\vb k)+B\delta N_{\delta X}(\vb k)+\order{B^2},
\end{equation*}
where
\begin{align}\label{Eq:dNdX}
	\delta N_{\delta X}(\vb k)=\lim_{B\to 0}\frac{f_{\text{FD}}[H_0^{\text{MF}}(\vb k)+B\delta X(\vb k)]-f_{\text{FD}}[H_0^{\text{MF}}(\vb k)]}B
\end{align}
can be calculated by regarding $B\delta X$ as a perturbation to $H_0^{\text{MF}}$ (see Appendix~\ref{App:density}).
Putting the above expansion of $X_B(\vb k)$ and $N_B(\vb k)$ back to Eq.~\eqref{Eq:SC0} and collecting the first order in $B$, we find that $\delta X$ satisfies the following linear equation
\begin{align}\label{Eq:deltaX}
	\delta X(\vb k)=V_{\text{MF}}[\delta N_{\pi}(\vb k)+\delta N_{\delta X}(\vb k)],
\end{align}
which can be efficiently solved numerically.
Here $\delta N_{\pi}$ is a fixed source determined entirely by the zero-field solution, whereas $\delta N_{\delta X}$ is linear in $\delta X$. 
We also note that in our convention, notations like $\delta{X}$ always represent a rate or change rather than an increment, and carry an extra dimension of of $1/B$ relative to $X$.

Moreover, using the analytic expressions in Appendix~\ref{App:density}, we also obtain $\pdv*{n}{B}$ in the HF approximation,
\begin{align}\label{Eq:dndB}
	\pdv{n}{B}&=\int\frac{\dd[2]{k}}{(2\pi)^2}\left\{\trace[\delta N_{\pi}(\vb k)]+\trace[\delta N_{\delta X}(\vb k)]\right\}\nonumber\\
	&=\sum_{\alpha}\int\frac{\dd[2]{k}}{(2\pi)^2}\qty(f_\alpha\Omega_\alpha-f_\alpha' m_\alpha+f_\alpha'\delta X_\alpha),
\end{align}
where the index $\alpha$ refers to $\ket{\alpha}$, the eigenstate of $H_0^{\text{MF}}$ with eigenvalue $\varepsilon_\alpha$, $f_\alpha'=f_{\text{FD}}'(\varepsilon_\alpha)$, $\Omega_\alpha=-2\text{Im}\braket{\partial_x\alpha}{\partial_y\alpha}$ is the Berry curvature, $m_\alpha=\text{Im}\mel{\partial_x\alpha}{(H_0^{\text{MF}}-\varepsilon_\alpha)}{\partial_y\alpha}$ is the orbital magnetic moment, and $\delta X_\alpha=\mel{\alpha}{\delta X}{\alpha}$ is the expectation value of $\delta X$ on $\ket{\alpha}$. The first two terms are the well-known result for noninteracting systems~\cite{XiaoShiNiu2005,XiaoChangNiu2010}, and the HF correction produces an additional Fermi-surface term. Nonetheless, the additional term, together with the contribution from orbital magnetic moment, vanishes for insulating states, leading to the \streda\ formula~\cite{Streda1982}. 
This proves that the HF approximation is compatible with the \streda\ formula for any insulating state, in agreement with previous studies~\cite{Huang2026,Zhu2026}.

\lettersection{Generalization to periodic systems}
The previous discussion assumed translation symmetry; we now generalize the results to generic periodic systems. 
For periodic systems, the single-particle Hamiltonian $\hat{h}$ has only lattice translation symmetry, that is, $[e^{i\vb R\vdot\hat{\vb Q}},\hat h]=0$ for all real-space lattice vectors $\vb R$; equivalently, the position dependence of $\hat h$ has Fourier support on the set $\Lambda$ of all reciprocal lattice vectors. 
Hence, neither $\hat{h}$ nor $\hat{H}^{\text{MF}}$ can be expressed as a function of $\hat{\vb*\pi}$. In fact, we prove in Appendix~\ref{App:peierls} that operators respecting lattice translation symmetry can be expressed as a function of $\hat{\vb*\pi}$ and $\hat{\vb r}$ with the following unique decomposition,
\begin{align}\label{Eq:decomp}
	A(\hat{\vb*\pi},\hat{\vb r})=\sum_{\vb g\in\Lambda}A_{\vb g}(\hat{\vb*\pi})e^{i\vb g\vdot\hat{\vb r}},
\end{align}
and the mean-field potential becomes
\begin{align}\label{Eq:HF}
	V_{\text{MF}}[N(\hat{\vb*\pi},\hat{\vb r})]&=\mathcal{A}^{-1}\sum_{\vb g\in\Lambda}\tilde V(g)\trace[ N_{\vb g}(\hat{\vb*\pi})]e^{i\vb g\vdot \hat{\vb r}}\nonumber\\
	&\quad-\int\frac{\dd[2]{q}}{(2\pi)^2}\tilde V(q)N(\hat{\vb*\pi}-\vb q,\hat{\vb r}),
\end{align}
where $\mathcal{A}$ is the area of the system.
The appearance of the position operator complicates the discussion, but it can be circumvented by extending the Hilbert space. 

\begin{figure*}[!t]
	\center
	\includegraphics[width=\textwidth]{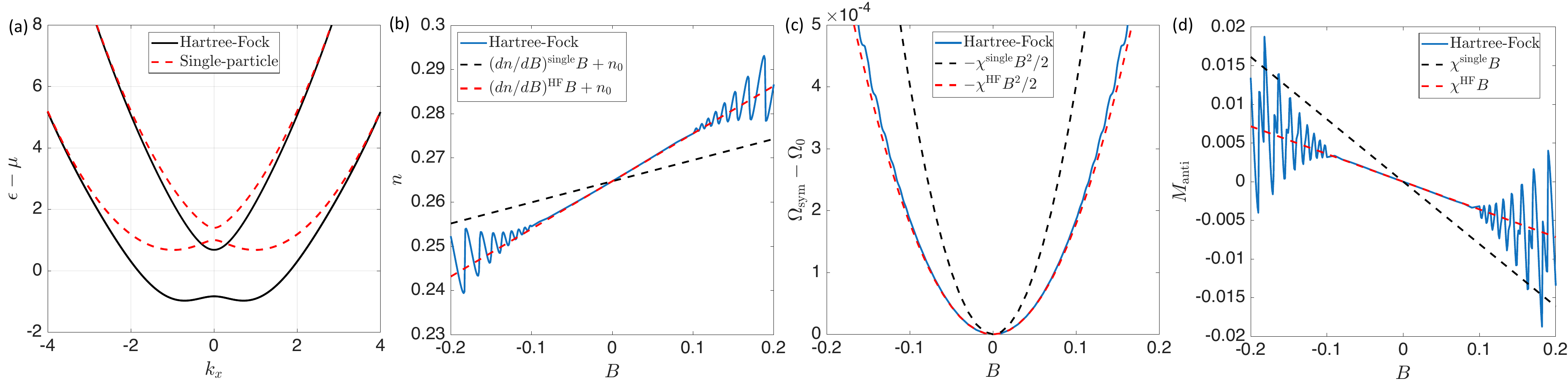}
	\caption{\label{Fig:TI}HF Calculation in the translation-invariant toy model. (a) HF and single-particle band structure at zero field. (b) Total density as a function of magnetic field obtained by finite-field HF calculation. 
	(c) HF grand potential density as a function of magnetic field. (d) HF magnetization as a function of magnetic field. The red dashed lines use the full HF formula, while the black dashed lines use the single-particle formula. Here, we perform the HF calculation at a small but finite temperature $T=0.01$ to suppress quantum oscillations.
	}
\end{figure*}

Operators respecting lattice translation symmetry clearly form an algebra, and any set of operators sharing the same algebra essentially describes the same physics. Inspired by this concept, we introduce an extended Hilbert space $\mathcal H_{k}\otimes\mathcal H'$, which is a tensor product of the Hilbert space for momentum $\mathcal H_{k}$ and an auxiliary Hilbert space $\mathcal H'$ spanned by orthogonal basis $\{\ket{\vb g}\}_{\vb g\in\Lambda}$. We map $\hat A$ in Eq.~\eqref{Eq:decomp} to
\begin{align}
	M(\hat{A}):=\sum_{\vb g,\vb g'\in\Lambda}A_{\vb g}(\hat{\vb*\pi}+\vb g')\dyad{\vb g'}{\vb g'-\vb g},
\end{align}
where $A_{\vb g}(\hat{\vb*\pi})$ are operators in $\mathcal H_{k}$.
It is easy to verify that this linear map is injective and preserves the operator algebra, as shown in Appendix~\ref{App:algebra}.
Therefore, any periodic operator can be mapped to a matrix-valued function of $\hat{\vb*\pi}$, and the noncommutativity between $e^{i\vb g\vdot\hat{\vb r}}$ and $\hat{\vb*\pi}$ is encoded in the noncommutativity between matrices. Hence, we can map the self-consistent equations to the extended Hilbert space so that the system becomes translation-invariant. Moreover, as only operator algebra is used in the previous section, the conclusion Eq.~\eqref{Eq:deltaX} and~\eqref{Eq:dndB} can be directly applied to the periodic case with the $V_{\text{MF}}$ given in Eq.~\eqref{Eq:HF}. 

In particular, $M(\hat h)$ at zero field is exactly the matrix used to solve the continuum models of \moire systems. The difference is that the momentum is only defined in the first Brillouin zone (BZ) in the physical Hilbert space, while the momentum is defined on the entire plane in the extended Hilbert space. Hence, the extension is equivalent to making an infinite number of copies of the first BZ. As a result, the operator algebra remains unchanged, but the system gains translational symmetry. $M(\hat h)$ at finite fields has also been explored as an alternative way to implement Peierls substitution~\cite{Lian2021}. By truncating $\mathcal H'$ at finite fields as one does at zero field, edge states emerge and reflect the topology of the system.

\lettersection{Magnetic susceptibility}
We now discuss how the HF correction $\delta X$ affects the magnetic susceptibility. Magnetic susceptibility is the rate of change of magnetization, and magnetization in the HF theory is known to have a form identical to the single-particle magnetization at both zero and finite fields~\cite{XiaoShiNiu2005,Thonhauser2005,ShiVignaleXiaoNiu2007,XiaoChangNiu2010}.
Moreover, the single-particle magnetization at finite fields can be expanded in terms of $B$,
\begin{align}
	M^{\text{single}}_{H(\hat{\vb*\pi})}=M^{\text{single}}_{H(\hat{\vb k})}+B\chi^{\text{single}}_{H(\hat{\vb k})}+\order{B^2},
\end{align}
where $M^{\text{single}}_{H(\hat{\vb*\pi})}$ and $M^{\text{single}}_{H(\hat{\vb k})}$ are respectively the single-particle magnetization of $H(\hat{\vb*\pi})$ and $H(\hat{\vb k})$, and $\chi^{\text{single}}_{H(\hat{\vb k})}$ is the single-particle susceptibility of $H(\hat{\vb k})$~\cite{Gao2015,Raoux2015}. Hence, the HF magnetization becomes
\begin{align}\label{Eq:HFMag}
	M^{\text{HF}}_B&=M^{\text{single}}_{H_B^{\text{MF}}(\hat{\vb*\pi})}=M^{\text{single}}_{H_B^{\text{MF}}(\hat{\vb k})}+B \chi^{\text{single}}_{H_B^{\text{MF}}(\hat{\vb k})}+\order{B^2}\nonumber\\
	&=M^{\text{single}}_{H_0^{\text{MF}}(\hat{\vb k})}+B\tilde\chi_{\delta X} +B \chi^{\text{single}}_{H_0^{\text{MF}}(\hat{\vb k})}+\order{B^2},
\end{align}
where $B\tilde\chi_{\delta X}$ describes the change of magnetization caused by perturbation $B\delta X$ and can be calculated by first-order perturbation theory (see Appendix~\ref{App:chidX}); in the last term $H_B^{\text{MF}}(\hat{\vb k})$ may be replaced by $H_0^{\text{MF}}(\hat{\vb k})$, as the correction to the subscript itself is of order $B$ and already multiplies an explicit factor of $B$. 
Hence, we obtain the central result of the work, 
\begin{align}\label{Eq:HFchi}
	\chi^{\text{HF}}=\lim_{B\to 0}\frac{M^{\text{HF}}_B-M^{\text{HF}}_0}{B}=\tilde\chi_{\delta X} +\chi^{\text{single}}_{H_0^{\text{MF}}(\hat{\vb k})}.
\end{align}
The additional term $\tilde\chi_{\delta X}$ can substantially change the susceptibility if $\delta X$ is considerably large. According to Eq.~\eqref{Eq:deltaX}, $\delta X$ is large when $\delta N_{\pi}$ is large. As $\trace[\delta N_{\pi}]\sim\sum_\alpha f_\alpha\Omega_\alpha$, we anticipate a large correction from this additional term if the occupied states contribute a large Berry curvature.

\begin{figure*}[!t]
	\center
	\includegraphics[width=\textwidth]{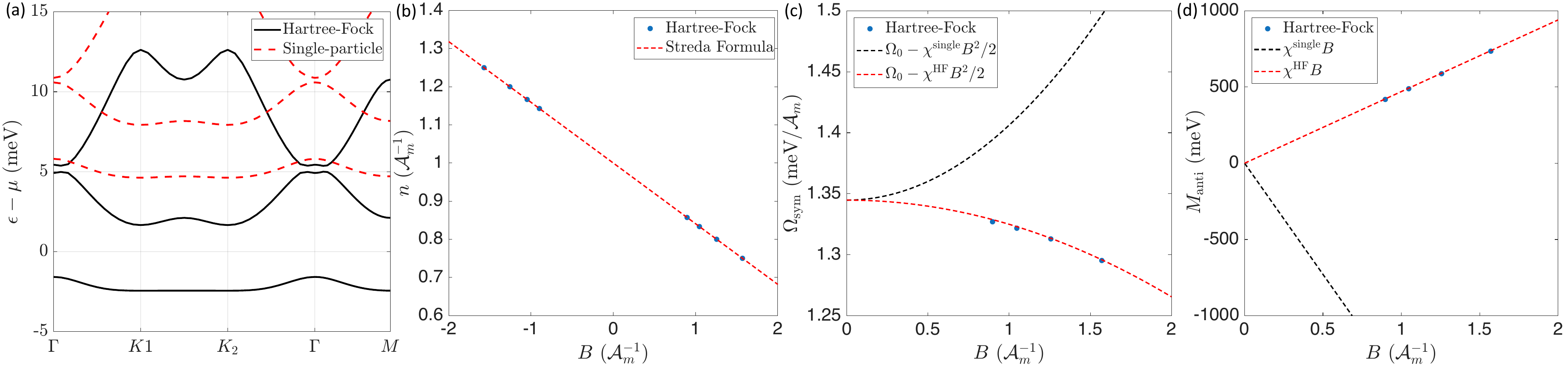}
	\caption{\label{Fig:moire}HF Calculation in the periodic toy model. (a) HF and single-particle band structure at zero field. (b) Total density as a function of magnetic field obtained by finite-field HF calculation. Here, $\mathcal{A}_m$ is the area of the \moire unit cell.
	(c) Symmetrized grand potential density $\Omega_{\text{sym}}$ as a function of magnetic field: the single-particle formula (black dashed) predicts the wrong curvature, i.e., the wrong sign of $\chi$. (d) Antisymmetrized magnetization: the finite-field HF data fall on the line of the full HF susceptibility $\chi^{\text{HF}}$ (red dashed), while the single-particle slope (black dashed) has the opposite sign.	}
\end{figure*}

\lettersection{Numerical results}
We numerically test our theory in both translation-invariant and periodic systems. We first study a translation-invariant single-particle Hamiltonian
\begin{align}\label{Eq:TI model}
	h(\vb k)=k^2/(2m)+v\vb k\vdot\vb*\sigma+\Delta\sigma_z
\end{align}
with a Gaussian interaction $\tilde V(q)=Ue^{-q^2/2\lambda^2}$. The parameters are given in Appendix~\ref{App:models}. In this system, there is a Dirac cone at $\vb k=0$ with a small gap opened by $\Delta$ and the interaction, as shown by the HF band structure at zero field in Fig.~\ref{Fig:TI}(a). The occupied states have a large Berry curvature and consequently a large $\delta X$, if only the lowest band at $\vb k=0$ is occupied. The expected correction from $\delta X$ is demonstrated in Fig.~\ref{Fig:TI}(b), where we perform HF calculations at finite fields with fixed chemical potential $\mu=-1.2$. As the ground state is metallic, we introduce a small temperature of $T=0.01$ to suppress quantum oscillations. The slope of the total density is accurately captured by Eq.~\eqref{Eq:dndB}, whereas the single-particle formula, the first two terms in Eq.~\eqref{Eq:dndB}, significantly deviates from the finite-field HF results. The indispensable role of $\delta X$ is also highlighted by its contribution to magnetic susceptibility. We numerically verify the magnetic susceptibility by calculating symmetrized grand potential density $\Omega_{\text{sym}}=[\Omega(B)+\Omega(-B)]/2=\Omega_0-\chi B^2/2+\order{B^4}$ in Fig.~\ref{Fig:TI}(c) and antisymmetrized magnetization $M_{\text{anti}}=[M(B)-M(-B)]/2=\chi B+\order{B^3}$ in Fig.~\ref{Fig:TI}(d). Both show striking agreement between our interacting theory of magnetic susceptibility and the numerical results at finite fields. Meanwhile, the single-particle formula $\chi^{\text{single}}_{H_0^{\text{MF}}(\hat{\vb k})}$ alone cannot correctly describe the system.

Second, we study a topologically nontrivial periodic system, the continuum model of tMoTe$_2$~\cite{Wu2019}, and perform unprojected HF calculations at both zero and finite fields. We only use it as a toy model and ignore the spin and valley degrees of freedom. Moreover, we adopt the hole language rather than the electron language, so the topmost valence band of Ref.~\cite{Wu2019} becomes the lowest band of Eq.~\eqref{Eq:moire model}, with a positive effective mass. The single-particle Hamiltonian is given by
\begin{align}\label{Eq:moire model}
	\hat h=\mqty[\hat k_t^2/(2m)+V_t(\hat{\vb r}) & T(\hat{\vb r}) \\ T(\hat{\vb r})^\dag & \hat k_b^2/(2m)+V_b(\hat{\vb r})],
\end{align}
where $\hat{\vb*k}_{t/b}=\hat{\vb*k}-\vb*\kappa_{t/b}$ is the momentum of top/bottom layer, $\vb*\kappa_{t/b}$ is the K valley of the two layers, $V_{t/b}(\hat{\vb r})$ describe the \moire potential within top and bottom MoTe$_2$, and $T(\hat{\vb r})$ describes the tunneling between the two layers.
Furthermore, we consider the dual-gate screened Coulomb potential $\tilde V(q)=e^2\tanh(qd)/(4\pi \epsilon \epsilon_0 q)$ with $d=\SI{30}{nm}$ and $\epsilon=25$. The other parameters and a detailed description of the model are given in Appendix~\ref{App:models}. 
The lowest band of the single-particle system has Chern number $C=-1$, and we fix $\mu=-\SI{23}{meV}$ so that the HF ground state at zero field is a Chern insulator, as shown by the HF band structure in Fig.~\ref{Fig:moire}(a). In Fig.~\ref{Fig:moire}(b), we calculate the total density of the HF ground state as a function of magnetic field, which exactly follows the \streda\ formula as we proved. 
We emphasize that in spite of the theoretical equivalence, the conventional Peierls substitution~\cite{Wang2024} is more suitable for numerical HF calculations than the open-space one~\cite{Lian2021}, because the conventional one has no edge states. 
Note that while $\delta X$ does not contribute to $\pdv*{n}{B}$ for insulators, it can still provide a substantial correction to the magnetic susceptibility. 
In Fig.~\ref{Fig:moire}(c), we calculate the symmetrized grand potential $\Omega_{\text{sym}}$, which agrees with $\chi^{\text{HF}}$ in Eq.~\eqref{Eq:HFchi} remarkably well. 
In contrast, $\chi^{\text{single}}_{H_0^{\text{MF}}(\hat{\vb k})}$ alone cannot even give the correct sign of the magnetic susceptibility. 
The accuracy of our theory is better presented by the antisymmetrized magnetization, which has better numerical precision than energy, because $\chi^{\text{HF}}$ is a second-order correction to energy but a first-order correction to magnetization. As shown in Fig.~\ref{Fig:moire}(d), the magnetization falls perfectly on the theoretical line of $\chi^{\text{HF}}$. By contrast, the single-particle formula $\chi^{\text{single}}_{H_0^{\text{MF}}(\hat{\vb k})}$ is far from the actual magnetization.

\lettersection{Summary and outlook}
We have developed an algebraic theory of the Hartree-Fock response to a magnetic field: the reverse Peierls substitution converts the finite-field self-consistency into a function equation continuously connected to zero field, whose $\order{B}$ expansion yields a linear equation for the field-induced Fock correction $\delta X$; periodic and \moire systems enter as translation-invariant ones in an extended Hilbert space. 
The tMoTe$_2$ calculation carries a direct experimental message: for interacting Chern bands, the single-particle susceptibility evaluated on the self-consistent bands can fail even in sign, so the interaction term $\tilde\chi_{\delta X}$ is indispensable for interpreting magnetic and thermodynamic measurements of \moire Chern insulators~\cite{Tschirhart2021,Zeng2023,Redekop2024}. 
The framework we developed is gauge invariant by construction and systematically extendable: Eq.~\eqref{Eq:deltaX} generalizes to arbitrary order in $B$, giving immediate access to nonlinear magnetic responses.

\lettersection{Acknowledgments}
This work is supported by the Research Grants Council of Hong Kong (Grants No. CityU 11304823, CityU 11312825, C7012-21G, and C7015-24G) and City University of Hong Kong (Project No. 9610428).

\bibliography{refs}

\begin{thebibliography}{36}%
\makeatletter
\providecommand \@ifxundefined [1]{%
 \@ifx{#1\undefined}
}%
\providecommand \@ifnum [1]{%
 \ifnum #1\expandafter \@firstoftwo
 \else \expandafter \@secondoftwo
 \fi
}%
\providecommand \@ifx [1]{%
 \ifx #1\expandafter \@firstoftwo
 \else \expandafter \@secondoftwo
 \fi
}%
\providecommand \natexlab [1]{#1}%
\providecommand \enquote  [1]{``#1''}%
\providecommand \bibnamefont  [1]{#1}%
\providecommand \bibfnamefont [1]{#1}%
\providecommand \citenamefont [1]{#1}%
\providecommand \href@noop [0]{\@secondoftwo}%
\providecommand \href [0]{\begingroup \@sanitize@url \@href}%
\providecommand \@href[1]{\@@startlink{#1}\@@href}%
\providecommand \@@href[1]{\endgroup#1\@@endlink}%
\providecommand \@sanitize@url [0]{\catcode `\\12\catcode `\$12\catcode
  `\&12\catcode `\#12\catcode `\^12\catcode `\_12\catcode `\%12\relax}%
\providecommand \@@startlink[1]{}%
\providecommand \@@endlink[0]{}%
\providecommand \url  [0]{\begingroup\@sanitize@url \@url }%
\providecommand \@url [1]{\endgroup\@href {#1}{\urlprefix }}%
\providecommand \urlprefix  [0]{URL }%
\providecommand \Eprint [0]{\href }%
\providecommand \doibase [0]{https://doi.org/}%
\providecommand \selectlanguage [0]{\@gobble}%
\providecommand \bibinfo  [0]{\@secondoftwo}%
\providecommand \bibfield  [0]{\@secondoftwo}%
\providecommand \translation [1]{[#1]}%
\providecommand \BibitemOpen [0]{}%
\providecommand \bibitemStop [0]{}%
\providecommand \bibitemNoStop [0]{.\EOS\space}%
\providecommand \EOS [0]{\spacefactor3000\relax}%
\providecommand \BibitemShut  [1]{\csname bibitem#1\endcsname}%
\let\auto@bib@innerbib\@empty
\bibitem [{\citenamefont {Finkler}\ \emph {et~al.}(2012)\citenamefont
  {Finkler}, \citenamefont {Vasyukov}, \citenamefont {Segev}, \citenamefont
  {Ne'eman}, \citenamefont {Lachman}, \citenamefont {Rappaport}, \citenamefont
  {Myasoedov}, \citenamefont {Zeldov},\ and\ \citenamefont
  {Huber}}]{Finkler2012}%
  \BibitemOpen
  \bibfield  {author} {\bibinfo {author} {\bibfnamefont {A.}~\bibnamefont
  {Finkler}}, \bibinfo {author} {\bibfnamefont {D.}~\bibnamefont {Vasyukov}},
  \bibinfo {author} {\bibfnamefont {Y.}~\bibnamefont {Segev}}, \bibinfo
  {author} {\bibfnamefont {L.}~\bibnamefont {Ne'eman}}, \bibinfo {author}
  {\bibfnamefont {E.~O.}\ \bibnamefont {Lachman}}, \bibinfo {author}
  {\bibfnamefont {M.~L.}\ \bibnamefont {Rappaport}}, \bibinfo {author}
  {\bibfnamefont {Y.}~\bibnamefont {Myasoedov}}, \bibinfo {author}
  {\bibfnamefont {E.}~\bibnamefont {Zeldov}},\ and\ \bibinfo {author}
  {\bibfnamefont {M.~E.}\ \bibnamefont {Huber}},\ }\bibfield  {title} {\bibinfo
  {title} {Scanning superconducting quantum interference device on a tip for
  magnetic imaging of nanoscale phenomena},\ }\href
  {https://doi.org/10.1063/1.4731656} {\bibfield  {journal} {\bibinfo
  {journal} {Rev. Sci. Instrum.}\ }\textbf {\bibinfo {volume} {83}},\ \bibinfo
  {pages} {073702} (\bibinfo {year} {2012})}\BibitemShut {NoStop}%
\bibitem [{\citenamefont {Vasyukov}\ \emph {et~al.}(2013)\citenamefont
  {Vasyukov}, \citenamefont {Anahory}, \citenamefont {Embon}, \citenamefont
  {Halbertal}, \citenamefont {Cuppens}, \citenamefont {Ne'eman}, \citenamefont
  {Finkler}, \citenamefont {Segev}, \citenamefont {Myasoedov}, \citenamefont
  {Rappaport}, \citenamefont {Huber},\ and\ \citenamefont
  {Zeldov}}]{Vasyukov2013}%
  \BibitemOpen
  \bibfield  {author} {\bibinfo {author} {\bibfnamefont {D.}~\bibnamefont
  {Vasyukov}}, \bibinfo {author} {\bibfnamefont {Y.}~\bibnamefont {Anahory}},
  \bibinfo {author} {\bibfnamefont {L.}~\bibnamefont {Embon}}, \bibinfo
  {author} {\bibfnamefont {D.}~\bibnamefont {Halbertal}}, \bibinfo {author}
  {\bibfnamefont {J.}~\bibnamefont {Cuppens}}, \bibinfo {author} {\bibfnamefont
  {L.}~\bibnamefont {Ne'eman}}, \bibinfo {author} {\bibfnamefont
  {A.}~\bibnamefont {Finkler}}, \bibinfo {author} {\bibfnamefont
  {Y.}~\bibnamefont {Segev}}, \bibinfo {author} {\bibfnamefont
  {Y.}~\bibnamefont {Myasoedov}}, \bibinfo {author} {\bibfnamefont {M.~L.}\
  \bibnamefont {Rappaport}}, \bibinfo {author} {\bibfnamefont {M.~E.}\
  \bibnamefont {Huber}},\ and\ \bibinfo {author} {\bibfnamefont
  {E.}~\bibnamefont {Zeldov}},\ }\bibfield  {title} {\bibinfo {title} {A
  scanning superconducting quantum interference device with single electron
  spin sensitivity},\ }\href {https://doi.org/10.1038/nnano.2013.169}
  {\bibfield  {journal} {\bibinfo  {journal} {Nat. Nanotechnol.}\ }\textbf
  {\bibinfo {volume} {8}},\ \bibinfo {pages} {639} (\bibinfo {year}
  {2013})}\BibitemShut {NoStop}%
\bibitem [{\citenamefont {Anahory}\ \emph {et~al.}(2020)\citenamefont
  {Anahory}, \citenamefont {Naren}, \citenamefont {Lachman}, \citenamefont
  {Buhbut~Sinai}, \citenamefont {Uri}, \citenamefont {Embon}, \citenamefont
  {Yaakobi}, \citenamefont {Myasoedov}, \citenamefont {Huber}, \citenamefont
  {Klajn},\ and\ \citenamefont {Zeldov}}]{Anahory2020}%
  \BibitemOpen
  \bibfield  {author} {\bibinfo {author} {\bibfnamefont {Y.}~\bibnamefont
  {Anahory}}, \bibinfo {author} {\bibfnamefont {H.~R.}\ \bibnamefont {Naren}},
  \bibinfo {author} {\bibfnamefont {E.~O.}\ \bibnamefont {Lachman}}, \bibinfo
  {author} {\bibfnamefont {S.}~\bibnamefont {Buhbut~Sinai}}, \bibinfo {author}
  {\bibfnamefont {A.}~\bibnamefont {Uri}}, \bibinfo {author} {\bibfnamefont
  {L.}~\bibnamefont {Embon}}, \bibinfo {author} {\bibfnamefont
  {E.}~\bibnamefont {Yaakobi}}, \bibinfo {author} {\bibfnamefont
  {Y.}~\bibnamefont {Myasoedov}}, \bibinfo {author} {\bibfnamefont {M.~E.}\
  \bibnamefont {Huber}}, \bibinfo {author} {\bibfnamefont {R.}~\bibnamefont
  {Klajn}},\ and\ \bibinfo {author} {\bibfnamefont {E.}~\bibnamefont
  {Zeldov}},\ }\bibfield  {title} {\bibinfo {title} {{SQUID}-on-tip with
  single-electron spin sensitivity for high-field and ultra-low temperature
  nanomagnetic imaging},\ }\href {https://doi.org/10.1039/C9NR08578E}
  {\bibfield  {journal} {\bibinfo  {journal} {Nanoscale}\ }\textbf {\bibinfo
  {volume} {12}},\ \bibinfo {pages} {3174} (\bibinfo {year}
  {2020})}\BibitemShut {NoStop}%
\bibitem [{\citenamefont {Zhou}\ \emph
  {et~al.}(2023{\natexlab{a}})\citenamefont {Zhou}, \citenamefont {Auerbach},
  \citenamefont {Roy}, \citenamefont {Bocarsly}, \citenamefont {Huber},
  \citenamefont {Barick}, \citenamefont {Pariari}, \citenamefont {H\"ucker},
  \citenamefont {Lim}, \citenamefont {Ariando}, \citenamefont {Berdyugin},
  \citenamefont {Xin}, \citenamefont {Rappaport}, \citenamefont {Myasoedov},\
  and\ \citenamefont {Zeldov}}]{ZhouSOT2023}%
  \BibitemOpen
  \bibfield  {author} {\bibinfo {author} {\bibfnamefont {H.}~\bibnamefont
  {Zhou}}, \bibinfo {author} {\bibfnamefont {N.}~\bibnamefont {Auerbach}},
  \bibinfo {author} {\bibfnamefont {I.}~\bibnamefont {Roy}}, \bibinfo {author}
  {\bibfnamefont {M.}~\bibnamefont {Bocarsly}}, \bibinfo {author}
  {\bibfnamefont {M.~E.}\ \bibnamefont {Huber}}, \bibinfo {author}
  {\bibfnamefont {B.}~\bibnamefont {Barick}}, \bibinfo {author} {\bibfnamefont
  {A.}~\bibnamefont {Pariari}}, \bibinfo {author} {\bibfnamefont
  {M.}~\bibnamefont {H\"ucker}}, \bibinfo {author} {\bibfnamefont {Z.~S.}\
  \bibnamefont {Lim}}, \bibinfo {author} {\bibfnamefont {A.}~\bibnamefont
  {Ariando}}, \bibinfo {author} {\bibfnamefont {A.~I.}\ \bibnamefont
  {Berdyugin}}, \bibinfo {author} {\bibfnamefont {N.}~\bibnamefont {Xin}},
  \bibinfo {author} {\bibfnamefont {M.}~\bibnamefont {Rappaport}}, \bibinfo
  {author} {\bibfnamefont {Y.}~\bibnamefont {Myasoedov}},\ and\ \bibinfo
  {author} {\bibfnamefont {E.}~\bibnamefont {Zeldov}},\ }\bibfield  {title}
  {\bibinfo {title} {Scanning {SQUID}-on-tip microscope in a top-loading
  cryogen-free dilution refrigerator},\ }\href
  {https://doi.org/10.1063/5.0142073} {\bibfield  {journal} {\bibinfo
  {journal} {Rev. Sci. Instrum.}\ }\textbf {\bibinfo {volume} {94}},\ \bibinfo
  {pages} {053706} (\bibinfo {year} {2023}{\natexlab{a}})}\BibitemShut
  {NoStop}%
\bibitem [{\citenamefont {Sharpe}\ \emph {et~al.}(2019)\citenamefont {Sharpe},
  \citenamefont {Fox}, \citenamefont {Barnard}, \citenamefont {Finney},
  \citenamefont {Watanabe}, \citenamefont {Taniguchi}, \citenamefont
  {Kastner},\ and\ \citenamefont {Goldhaber-Gordon}}]{Sharpe2019}%
  \BibitemOpen
  \bibfield  {author} {\bibinfo {author} {\bibfnamefont {A.~L.}\ \bibnamefont
  {Sharpe}}, \bibinfo {author} {\bibfnamefont {E.~J.}\ \bibnamefont {Fox}},
  \bibinfo {author} {\bibfnamefont {A.~W.}\ \bibnamefont {Barnard}}, \bibinfo
  {author} {\bibfnamefont {J.}~\bibnamefont {Finney}}, \bibinfo {author}
  {\bibfnamefont {K.}~\bibnamefont {Watanabe}}, \bibinfo {author}
  {\bibfnamefont {T.}~\bibnamefont {Taniguchi}}, \bibinfo {author}
  {\bibfnamefont {M.~A.}\ \bibnamefont {Kastner}},\ and\ \bibinfo {author}
  {\bibfnamefont {D.}~\bibnamefont {Goldhaber-Gordon}},\ }\bibfield  {title}
  {\bibinfo {title} {Emergent ferromagnetism near three-quarters filling in
  twisted bilayer graphene},\ }\href {https://doi.org/10.1126/science.aaw3780}
  {\bibfield  {journal} {\bibinfo  {journal} {Science}\ }\textbf {\bibinfo
  {volume} {365}},\ \bibinfo {pages} {605} (\bibinfo {year}
  {2019})}\BibitemShut {NoStop}%
\bibitem [{\citenamefont {Serlin}\ \emph {et~al.}(2020)\citenamefont {Serlin},
  \citenamefont {Tschirhart}, \citenamefont {Polshyn}, \citenamefont {Zhang},
  \citenamefont {Zhu}, \citenamefont {Watanabe}, \citenamefont {Taniguchi},
  \citenamefont {Balents},\ and\ \citenamefont {Young}}]{Serlin2020}%
  \BibitemOpen
  \bibfield  {author} {\bibinfo {author} {\bibfnamefont {M.}~\bibnamefont
  {Serlin}}, \bibinfo {author} {\bibfnamefont {C.~L.}\ \bibnamefont
  {Tschirhart}}, \bibinfo {author} {\bibfnamefont {H.}~\bibnamefont {Polshyn}},
  \bibinfo {author} {\bibfnamefont {Y.}~\bibnamefont {Zhang}}, \bibinfo
  {author} {\bibfnamefont {J.}~\bibnamefont {Zhu}}, \bibinfo {author}
  {\bibfnamefont {K.}~\bibnamefont {Watanabe}}, \bibinfo {author}
  {\bibfnamefont {T.}~\bibnamefont {Taniguchi}}, \bibinfo {author}
  {\bibfnamefont {L.}~\bibnamefont {Balents}},\ and\ \bibinfo {author}
  {\bibfnamefont {A.~F.}\ \bibnamefont {Young}},\ }\bibfield  {title} {\bibinfo
  {title} {Intrinsic quantized anomalous {H}all effect in a moir\'e
  heterostructure},\ }\href {https://doi.org/10.1126/science.aay5533}
  {\bibfield  {journal} {\bibinfo  {journal} {Science}\ }\textbf {\bibinfo
  {volume} {367}},\ \bibinfo {pages} {900} (\bibinfo {year}
  {2020})}\BibitemShut {NoStop}%
\bibitem [{\citenamefont {Cai}\ \emph {et~al.}(2023)\citenamefont {Cai},
  \citenamefont {Anderson}, \citenamefont {Wang}, \citenamefont {Zhang},
  \citenamefont {Liu}, \citenamefont {Holtzmann}, \citenamefont {Zhang},
  \citenamefont {Fan}, \citenamefont {Taniguchi}, \citenamefont {Watanabe},
  \citenamefont {Ran}, \citenamefont {Cao}, \citenamefont {Fu}, \citenamefont
  {Xiao}, \citenamefont {Yao},\ and\ \citenamefont {Xu}}]{Cai2023}%
  \BibitemOpen
  \bibfield  {author} {\bibinfo {author} {\bibfnamefont {J.}~\bibnamefont
  {Cai}}, \bibinfo {author} {\bibfnamefont {E.}~\bibnamefont {Anderson}},
  \bibinfo {author} {\bibfnamefont {C.}~\bibnamefont {Wang}}, \bibinfo {author}
  {\bibfnamefont {X.}~\bibnamefont {Zhang}}, \bibinfo {author} {\bibfnamefont
  {X.}~\bibnamefont {Liu}}, \bibinfo {author} {\bibfnamefont {W.}~\bibnamefont
  {Holtzmann}}, \bibinfo {author} {\bibfnamefont {Y.}~\bibnamefont {Zhang}},
  \bibinfo {author} {\bibfnamefont {F.}~\bibnamefont {Fan}}, \bibinfo {author}
  {\bibfnamefont {T.}~\bibnamefont {Taniguchi}}, \bibinfo {author}
  {\bibfnamefont {K.}~\bibnamefont {Watanabe}}, \bibinfo {author}
  {\bibfnamefont {Y.}~\bibnamefont {Ran}}, \bibinfo {author} {\bibfnamefont
  {T.}~\bibnamefont {Cao}}, \bibinfo {author} {\bibfnamefont {L.}~\bibnamefont
  {Fu}}, \bibinfo {author} {\bibfnamefont {D.}~\bibnamefont {Xiao}}, \bibinfo
  {author} {\bibfnamefont {W.}~\bibnamefont {Yao}},\ and\ \bibinfo {author}
  {\bibfnamefont {X.}~\bibnamefont {Xu}},\ }\bibfield  {title} {\bibinfo
  {title} {Signatures of fractional quantum anomalous {H}all states in twisted
  {MoTe$_2$}},\ }\href {https://doi.org/10.1038/s41586-023-06289-w} {\bibfield
  {journal} {\bibinfo  {journal} {Nature}\ }\textbf {\bibinfo {volume} {622}},\
  \bibinfo {pages} {63} (\bibinfo {year} {2023})}\BibitemShut {NoStop}%
\bibitem [{\citenamefont {Zeng}\ \emph {et~al.}(2023)\citenamefont {Zeng},
  \citenamefont {Xia}, \citenamefont {Kang}, \citenamefont {Zhu}, \citenamefont
  {Kn\"uppel}, \citenamefont {Vaswani}, \citenamefont {Watanabe}, \citenamefont
  {Taniguchi}, \citenamefont {Mak},\ and\ \citenamefont {Shan}}]{Zeng2023}%
  \BibitemOpen
  \bibfield  {author} {\bibinfo {author} {\bibfnamefont {Y.}~\bibnamefont
  {Zeng}}, \bibinfo {author} {\bibfnamefont {Z.}~\bibnamefont {Xia}}, \bibinfo
  {author} {\bibfnamefont {K.}~\bibnamefont {Kang}}, \bibinfo {author}
  {\bibfnamefont {J.}~\bibnamefont {Zhu}}, \bibinfo {author} {\bibfnamefont
  {P.}~\bibnamefont {Kn\"uppel}}, \bibinfo {author} {\bibfnamefont
  {C.}~\bibnamefont {Vaswani}}, \bibinfo {author} {\bibfnamefont
  {K.}~\bibnamefont {Watanabe}}, \bibinfo {author} {\bibfnamefont
  {T.}~\bibnamefont {Taniguchi}}, \bibinfo {author} {\bibfnamefont {K.~F.}\
  \bibnamefont {Mak}},\ and\ \bibinfo {author} {\bibfnamefont {J.}~\bibnamefont
  {Shan}},\ }\bibfield  {title} {\bibinfo {title} {Thermodynamic evidence of
  fractional {C}hern insulator in moir\'e {MoTe$_2$}},\ }\href
  {https://doi.org/10.1038/s41586-023-06452-3} {\bibfield  {journal} {\bibinfo
  {journal} {Nature}\ }\textbf {\bibinfo {volume} {622}},\ \bibinfo {pages}
  {69} (\bibinfo {year} {2023})}\BibitemShut {NoStop}%
\bibitem [{\citenamefont {Park}\ \emph {et~al.}(2023)\citenamefont {Park},
  \citenamefont {Cai}, \citenamefont {Anderson}, \citenamefont {Zhang},
  \citenamefont {Zhu}, \citenamefont {Liu}, \citenamefont {Wang}, \citenamefont
  {Holtzmann}, \citenamefont {Hu}, \citenamefont {Liu}, \citenamefont
  {Taniguchi}, \citenamefont {Watanabe}, \citenamefont {Chu}, \citenamefont
  {Cao}, \citenamefont {Fu}, \citenamefont {Yao}, \citenamefont {Chang},
  \citenamefont {Cobden}, \citenamefont {Xiao},\ and\ \citenamefont
  {Xu}}]{Park2023}%
  \BibitemOpen
  \bibfield  {author} {\bibinfo {author} {\bibfnamefont {H.}~\bibnamefont
  {Park}}, \bibinfo {author} {\bibfnamefont {J.}~\bibnamefont {Cai}}, \bibinfo
  {author} {\bibfnamefont {E.}~\bibnamefont {Anderson}}, \bibinfo {author}
  {\bibfnamefont {Y.}~\bibnamefont {Zhang}}, \bibinfo {author} {\bibfnamefont
  {J.}~\bibnamefont {Zhu}}, \bibinfo {author} {\bibfnamefont {X.}~\bibnamefont
  {Liu}}, \bibinfo {author} {\bibfnamefont {C.}~\bibnamefont {Wang}}, \bibinfo
  {author} {\bibfnamefont {W.}~\bibnamefont {Holtzmann}}, \bibinfo {author}
  {\bibfnamefont {C.}~\bibnamefont {Hu}}, \bibinfo {author} {\bibfnamefont
  {Z.}~\bibnamefont {Liu}}, \bibinfo {author} {\bibfnamefont {T.}~\bibnamefont
  {Taniguchi}}, \bibinfo {author} {\bibfnamefont {K.}~\bibnamefont {Watanabe}},
  \bibinfo {author} {\bibfnamefont {J.-H.}\ \bibnamefont {Chu}}, \bibinfo
  {author} {\bibfnamefont {T.}~\bibnamefont {Cao}}, \bibinfo {author}
  {\bibfnamefont {L.}~\bibnamefont {Fu}}, \bibinfo {author} {\bibfnamefont
  {W.}~\bibnamefont {Yao}}, \bibinfo {author} {\bibfnamefont {C.-Z.}\
  \bibnamefont {Chang}}, \bibinfo {author} {\bibfnamefont {D.}~\bibnamefont
  {Cobden}}, \bibinfo {author} {\bibfnamefont {D.}~\bibnamefont {Xiao}},\ and\
  \bibinfo {author} {\bibfnamefont {X.}~\bibnamefont {Xu}},\ }\bibfield
  {title} {\bibinfo {title} {Observation of fractionally quantized anomalous
  {H}all effect},\ }\href {https://doi.org/10.1038/s41586-023-06536-0}
  {\bibfield  {journal} {\bibinfo  {journal} {Nature}\ }\textbf {\bibinfo
  {volume} {622}},\ \bibinfo {pages} {74} (\bibinfo {year} {2023})}\BibitemShut
  {NoStop}%
\bibitem [{\citenamefont {Xu}\ \emph {et~al.}(2023)\citenamefont {Xu},
  \citenamefont {Sun}, \citenamefont {Jia}, \citenamefont {Liu}, \citenamefont
  {Xu}, \citenamefont {Li}, \citenamefont {Gu}, \citenamefont {Watanabe},
  \citenamefont {Taniguchi}, \citenamefont {Tong}, \citenamefont {Jia},
  \citenamefont {Shi}, \citenamefont {Jiang}, \citenamefont {Zhang},
  \citenamefont {Liu},\ and\ \citenamefont {Li}}]{Xu2023}%
  \BibitemOpen
  \bibfield  {author} {\bibinfo {author} {\bibfnamefont {F.}~\bibnamefont
  {Xu}}, \bibinfo {author} {\bibfnamefont {Z.}~\bibnamefont {Sun}}, \bibinfo
  {author} {\bibfnamefont {T.}~\bibnamefont {Jia}}, \bibinfo {author}
  {\bibfnamefont {C.}~\bibnamefont {Liu}}, \bibinfo {author} {\bibfnamefont
  {C.}~\bibnamefont {Xu}}, \bibinfo {author} {\bibfnamefont {C.}~\bibnamefont
  {Li}}, \bibinfo {author} {\bibfnamefont {Y.}~\bibnamefont {Gu}}, \bibinfo
  {author} {\bibfnamefont {K.}~\bibnamefont {Watanabe}}, \bibinfo {author}
  {\bibfnamefont {T.}~\bibnamefont {Taniguchi}}, \bibinfo {author}
  {\bibfnamefont {B.}~\bibnamefont {Tong}}, \bibinfo {author} {\bibfnamefont
  {J.}~\bibnamefont {Jia}}, \bibinfo {author} {\bibfnamefont {Z.}~\bibnamefont
  {Shi}}, \bibinfo {author} {\bibfnamefont {S.}~\bibnamefont {Jiang}}, \bibinfo
  {author} {\bibfnamefont {Y.}~\bibnamefont {Zhang}}, \bibinfo {author}
  {\bibfnamefont {X.}~\bibnamefont {Liu}},\ and\ \bibinfo {author}
  {\bibfnamefont {T.}~\bibnamefont {Li}},\ }\bibfield  {title} {\bibinfo
  {title} {Observation of integer and fractional quantum anomalous {H}all
  effects in twisted bilayer {MoTe$_2$}},\ }\href
  {https://doi.org/10.1103/PhysRevX.13.031037} {\bibfield  {journal} {\bibinfo
  {journal} {Phys. Rev. X}\ }\textbf {\bibinfo {volume} {13}},\ \bibinfo
  {pages} {031037} (\bibinfo {year} {2023})}\BibitemShut {NoStop}%
\bibitem [{\citenamefont {Tschirhart}\ \emph {et~al.}(2021)\citenamefont
  {Tschirhart}, \citenamefont {Serlin}, \citenamefont {Polshyn}, \citenamefont
  {Shragai}, \citenamefont {Xia}, \citenamefont {Zhu}, \citenamefont {Zhang},
  \citenamefont {Watanabe}, \citenamefont {Taniguchi}, \citenamefont {Huber},\
  and\ \citenamefont {Young}}]{Tschirhart2021}%
  \BibitemOpen
  \bibfield  {author} {\bibinfo {author} {\bibfnamefont {C.~L.}\ \bibnamefont
  {Tschirhart}}, \bibinfo {author} {\bibfnamefont {M.}~\bibnamefont {Serlin}},
  \bibinfo {author} {\bibfnamefont {H.}~\bibnamefont {Polshyn}}, \bibinfo
  {author} {\bibfnamefont {A.}~\bibnamefont {Shragai}}, \bibinfo {author}
  {\bibfnamefont {Z.}~\bibnamefont {Xia}}, \bibinfo {author} {\bibfnamefont
  {J.}~\bibnamefont {Zhu}}, \bibinfo {author} {\bibfnamefont {Y.}~\bibnamefont
  {Zhang}}, \bibinfo {author} {\bibfnamefont {K.}~\bibnamefont {Watanabe}},
  \bibinfo {author} {\bibfnamefont {T.}~\bibnamefont {Taniguchi}}, \bibinfo
  {author} {\bibfnamefont {M.~E.}\ \bibnamefont {Huber}},\ and\ \bibinfo
  {author} {\bibfnamefont {A.~F.}\ \bibnamefont {Young}},\ }\bibfield  {title}
  {\bibinfo {title} {Imaging orbital ferromagnetism in a moir\'e {C}hern
  insulator},\ }\href {https://doi.org/10.1126/science.abd3190} {\bibfield
  {journal} {\bibinfo  {journal} {Science}\ }\textbf {\bibinfo {volume}
  {372}},\ \bibinfo {pages} {1323} (\bibinfo {year} {2021})}\BibitemShut
  {NoStop}%
\bibitem [{\citenamefont {Grover}\ \emph {et~al.}(2022)\citenamefont {Grover},
  \citenamefont {Bocarsly}, \citenamefont {Uri}, \citenamefont {Stepanov},
  \citenamefont {Di~Battista}, \citenamefont {Roy}, \citenamefont {Xiao},
  \citenamefont {Meltzer}, \citenamefont {Myasoedov}, \citenamefont {Pareek},
  \citenamefont {Watanabe}, \citenamefont {Taniguchi}, \citenamefont {Yan},
  \citenamefont {Stern}, \citenamefont {Berg}, \citenamefont {Efetov},\ and\
  \citenamefont {Zeldov}}]{Grover2022}%
  \BibitemOpen
  \bibfield  {author} {\bibinfo {author} {\bibfnamefont {S.}~\bibnamefont
  {Grover}}, \bibinfo {author} {\bibfnamefont {M.}~\bibnamefont {Bocarsly}},
  \bibinfo {author} {\bibfnamefont {A.}~\bibnamefont {Uri}}, \bibinfo {author}
  {\bibfnamefont {P.}~\bibnamefont {Stepanov}}, \bibinfo {author}
  {\bibfnamefont {G.}~\bibnamefont {Di~Battista}}, \bibinfo {author}
  {\bibfnamefont {I.}~\bibnamefont {Roy}}, \bibinfo {author} {\bibfnamefont
  {J.}~\bibnamefont {Xiao}}, \bibinfo {author} {\bibfnamefont {A.~Y.}\
  \bibnamefont {Meltzer}}, \bibinfo {author} {\bibfnamefont {Y.}~\bibnamefont
  {Myasoedov}}, \bibinfo {author} {\bibfnamefont {K.}~\bibnamefont {Pareek}},
  \bibinfo {author} {\bibfnamefont {K.}~\bibnamefont {Watanabe}}, \bibinfo
  {author} {\bibfnamefont {T.}~\bibnamefont {Taniguchi}}, \bibinfo {author}
  {\bibfnamefont {B.}~\bibnamefont {Yan}}, \bibinfo {author} {\bibfnamefont
  {A.}~\bibnamefont {Stern}}, \bibinfo {author} {\bibfnamefont
  {E.}~\bibnamefont {Berg}}, \bibinfo {author} {\bibfnamefont {D.~K.}\
  \bibnamefont {Efetov}},\ and\ \bibinfo {author} {\bibfnamefont
  {E.}~\bibnamefont {Zeldov}},\ }\bibfield  {title} {\bibinfo {title} {Chern
  mosaic and {B}erry-curvature magnetism in magic-angle graphene},\ }\href
  {https://doi.org/10.1038/s41567-022-01635-7} {\bibfield  {journal} {\bibinfo
  {journal} {Nat. Phys.}\ }\textbf {\bibinfo {volume} {18}},\ \bibinfo {pages}
  {885} (\bibinfo {year} {2022})}\BibitemShut {NoStop}%
\bibitem [{\citenamefont {Redekop}\ \emph {et~al.}(2024)\citenamefont
  {Redekop}, \citenamefont {Zhang}, \citenamefont {Park}, \citenamefont {Cai},
  \citenamefont {Anderson}, \citenamefont {Sheekey}, \citenamefont {Arp},
  \citenamefont {Babikyan}, \citenamefont {Salters}, \citenamefont {Watanabe},
  \citenamefont {Taniguchi}, \citenamefont {Huber}, \citenamefont {Xu},\ and\
  \citenamefont {Young}}]{Redekop2024}%
  \BibitemOpen
  \bibfield  {author} {\bibinfo {author} {\bibfnamefont {E.}~\bibnamefont
  {Redekop}}, \bibinfo {author} {\bibfnamefont {C.}~\bibnamefont {Zhang}},
  \bibinfo {author} {\bibfnamefont {H.}~\bibnamefont {Park}}, \bibinfo {author}
  {\bibfnamefont {J.}~\bibnamefont {Cai}}, \bibinfo {author} {\bibfnamefont
  {E.}~\bibnamefont {Anderson}}, \bibinfo {author} {\bibfnamefont
  {O.}~\bibnamefont {Sheekey}}, \bibinfo {author} {\bibfnamefont
  {T.}~\bibnamefont {Arp}}, \bibinfo {author} {\bibfnamefont {G.}~\bibnamefont
  {Babikyan}}, \bibinfo {author} {\bibfnamefont {S.}~\bibnamefont {Salters}},
  \bibinfo {author} {\bibfnamefont {K.}~\bibnamefont {Watanabe}}, \bibinfo
  {author} {\bibfnamefont {T.}~\bibnamefont {Taniguchi}}, \bibinfo {author}
  {\bibfnamefont {M.~E.}\ \bibnamefont {Huber}}, \bibinfo {author}
  {\bibfnamefont {X.}~\bibnamefont {Xu}},\ and\ \bibinfo {author}
  {\bibfnamefont {A.~F.}\ \bibnamefont {Young}},\ }\bibfield  {title} {\bibinfo
  {title} {Direct magnetic imaging of fractional {C}hern insulators in twisted
  {MoTe$_2$}},\ }\href {https://doi.org/10.1038/s41586-024-08153-x} {\bibfield
  {journal} {\bibinfo  {journal} {Nature}\ }\textbf {\bibinfo {volume} {635}},\
  \bibinfo {pages} {584} (\bibinfo {year} {2024})}\BibitemShut {NoStop}%
\bibitem [{\citenamefont {Auerbach}\ \emph {et~al.}(2025)\citenamefont
  {Auerbach}, \citenamefont {Dutta}, \citenamefont {Uzan}, \citenamefont
  {Vituri}, \citenamefont {Zhou}, \citenamefont {Meltzer}, \citenamefont
  {Grover}, \citenamefont {Holder}, \citenamefont {Emanuel}, \citenamefont
  {Huber}, \citenamefont {Myasoedov}, \citenamefont {Watanabe}, \citenamefont
  {Taniguchi}, \citenamefont {Oreg}, \citenamefont {Berg},\ and\ \citenamefont
  {Zeldov}}]{Auerbach2025}%
  \BibitemOpen
  \bibfield  {author} {\bibinfo {author} {\bibfnamefont {N.}~\bibnamefont
  {Auerbach}}, \bibinfo {author} {\bibfnamefont {S.}~\bibnamefont {Dutta}},
  \bibinfo {author} {\bibfnamefont {M.}~\bibnamefont {Uzan}}, \bibinfo {author}
  {\bibfnamefont {Y.}~\bibnamefont {Vituri}}, \bibinfo {author} {\bibfnamefont
  {Y.}~\bibnamefont {Zhou}}, \bibinfo {author} {\bibfnamefont {A.~Y.}\
  \bibnamefont {Meltzer}}, \bibinfo {author} {\bibfnamefont {S.}~\bibnamefont
  {Grover}}, \bibinfo {author} {\bibfnamefont {T.}~\bibnamefont {Holder}},
  \bibinfo {author} {\bibfnamefont {P.}~\bibnamefont {Emanuel}}, \bibinfo
  {author} {\bibfnamefont {M.~E.}\ \bibnamefont {Huber}}, \bibinfo {author}
  {\bibfnamefont {Y.}~\bibnamefont {Myasoedov}}, \bibinfo {author}
  {\bibfnamefont {K.}~\bibnamefont {Watanabe}}, \bibinfo {author}
  {\bibfnamefont {T.}~\bibnamefont {Taniguchi}}, \bibinfo {author}
  {\bibfnamefont {Y.}~\bibnamefont {Oreg}}, \bibinfo {author} {\bibfnamefont
  {E.}~\bibnamefont {Berg}},\ and\ \bibinfo {author} {\bibfnamefont
  {E.}~\bibnamefont {Zeldov}},\ }\bibfield  {title} {\bibinfo {title} {Isospin
  magnetic texture and intervalley exchange interaction in rhombohedral
  tetralayer graphene},\ }\href {https://doi.org/10.1038/s41567-025-03035-z}
  {\bibfield  {journal} {\bibinfo  {journal} {Nat. Phys.}\ }\textbf {\bibinfo
  {volume} {21}},\ \bibinfo {pages} {1765} (\bibinfo {year}
  {2025})}\BibitemShut {NoStop}%
\bibitem [{\citenamefont {Zhou}\ \emph
  {et~al.}(2023{\natexlab{b}})\citenamefont {Zhou}, \citenamefont {Auerbach},
  \citenamefont {Uzan}, \citenamefont {Zhou}, \citenamefont {Banu},
  \citenamefont {Zhi}, \citenamefont {Huber}, \citenamefont {Watanabe},
  \citenamefont {Taniguchi}, \citenamefont {Myasoedov}, \citenamefont {Yan},\
  and\ \citenamefont {Zeldov}}]{ZhouPMF2023}%
  \BibitemOpen
  \bibfield  {author} {\bibinfo {author} {\bibfnamefont {H.}~\bibnamefont
  {Zhou}}, \bibinfo {author} {\bibfnamefont {N.}~\bibnamefont {Auerbach}},
  \bibinfo {author} {\bibfnamefont {M.}~\bibnamefont {Uzan}}, \bibinfo {author}
  {\bibfnamefont {Y.}~\bibnamefont {Zhou}}, \bibinfo {author} {\bibfnamefont
  {N.}~\bibnamefont {Banu}}, \bibinfo {author} {\bibfnamefont {W.}~\bibnamefont
  {Zhi}}, \bibinfo {author} {\bibfnamefont {M.~E.}\ \bibnamefont {Huber}},
  \bibinfo {author} {\bibfnamefont {K.}~\bibnamefont {Watanabe}}, \bibinfo
  {author} {\bibfnamefont {T.}~\bibnamefont {Taniguchi}}, \bibinfo {author}
  {\bibfnamefont {Y.}~\bibnamefont {Myasoedov}}, \bibinfo {author}
  {\bibfnamefont {B.}~\bibnamefont {Yan}},\ and\ \bibinfo {author}
  {\bibfnamefont {E.}~\bibnamefont {Zeldov}},\ }\bibfield  {title} {\bibinfo
  {title} {Imaging quantum oscillations and millitesla pseudomagnetic fields in
  graphene},\ }\href {https://doi.org/10.1038/s41586-023-06763-5} {\bibfield
  {journal} {\bibinfo  {journal} {Nature}\ }\textbf {\bibinfo {volume} {624}},\
  \bibinfo {pages} {275} (\bibinfo {year} {2023}{\natexlab{b}})}\BibitemShut
  {NoStop}%
\bibitem [{\citenamefont {Shi}\ \emph {et~al.}(2007)\citenamefont {Shi},
  \citenamefont {Vignale}, \citenamefont {Xiao},\ and\ \citenamefont
  {Niu}}]{ShiVignaleXiaoNiu2007}%
  \BibitemOpen
  \bibfield  {author} {\bibinfo {author} {\bibfnamefont {J.}~\bibnamefont
  {Shi}}, \bibinfo {author} {\bibfnamefont {G.}~\bibnamefont {Vignale}},
  \bibinfo {author} {\bibfnamefont {D.}~\bibnamefont {Xiao}},\ and\ \bibinfo
  {author} {\bibfnamefont {Q.}~\bibnamefont {Niu}},\ }\bibfield  {title}
  {\bibinfo {title} {Quantum theory of orbital magnetization and its
  generalization to interacting systems},\ }\href
  {https://doi.org/10.1103/PhysRevLett.99.197202} {\bibfield  {journal}
  {\bibinfo  {journal} {Phys. Rev. Lett.}\ }\textbf {\bibinfo {volume} {99}},\
  \bibinfo {pages} {197202} (\bibinfo {year} {2007})}\BibitemShut {NoStop}%
\bibitem [{\citenamefont {Vignale}\ \emph {et~al.}(2026)\citenamefont
  {Vignale}, \citenamefont {Shi}, \citenamefont {Xiao},\ and\ \citenamefont
  {Niu}}]{Vignale2026}%
  \BibitemOpen
  \bibfield  {author} {\bibinfo {author} {\bibfnamefont {G.}~\bibnamefont
  {Vignale}}, \bibinfo {author} {\bibfnamefont {J.}~\bibnamefont {Shi}},
  \bibinfo {author} {\bibfnamefont {D.}~\bibnamefont {Xiao}},\ and\ \bibinfo
  {author} {\bibfnamefont {Q.}~\bibnamefont {Niu}},\ }\href@noop {} {\bibinfo
  {title} {Ward identities and orbital magnetization in current density
  functional theory}} (\bibinfo {year} {2026}),\ \Eprint
  {https://arxiv.org/abs/2605.14336} {arXiv:2605.14336} \BibitemShut {NoStop}%
\bibitem [{\citenamefont {Blount}(1962)}]{Blount1962}%
  \BibitemOpen
  \bibfield  {author} {\bibinfo {author} {\bibfnamefont {E.~I.}\ \bibnamefont
  {Blount}},\ }\bibfield  {title} {\bibinfo {title} {{B}loch electrons in a
  magnetic field},\ }\href {https://doi.org/10.1103/PhysRev.126.1636}
  {\bibfield  {journal} {\bibinfo  {journal} {Phys. Rev.}\ }\textbf {\bibinfo
  {volume} {126}},\ \bibinfo {pages} {1636} (\bibinfo {year}
  {1962})}\BibitemShut {NoStop}%
\bibitem [{\citenamefont {Roth}(1966)}]{Roth1966}%
  \BibitemOpen
  \bibfield  {author} {\bibinfo {author} {\bibfnamefont {L.~M.}\ \bibnamefont
  {Roth}},\ }\bibfield  {title} {\bibinfo {title} {Semiclassical theory of
  magnetic energy levels and magnetic susceptibility of {B}loch electrons},\
  }\href {https://doi.org/10.1103/PhysRev.145.434} {\bibfield  {journal}
  {\bibinfo  {journal} {Phys. Rev.}\ }\textbf {\bibinfo {volume} {145}},\
  \bibinfo {pages} {434} (\bibinfo {year} {1966})}\BibitemShut {NoStop}%
\bibitem [{\citenamefont {Xiao}\ \emph {et~al.}(2005)\citenamefont {Xiao},
  \citenamefont {Shi},\ and\ \citenamefont {Niu}}]{XiaoShiNiu2005}%
  \BibitemOpen
  \bibfield  {author} {\bibinfo {author} {\bibfnamefont {D.}~\bibnamefont
  {Xiao}}, \bibinfo {author} {\bibfnamefont {J.}~\bibnamefont {Shi}},\ and\
  \bibinfo {author} {\bibfnamefont {Q.}~\bibnamefont {Niu}},\ }\bibfield
  {title} {\bibinfo {title} {Berry phase correction to electron density of
  states in solids},\ }\href {https://doi.org/10.1103/PhysRevLett.95.137204}
  {\bibfield  {journal} {\bibinfo  {journal} {Phys. Rev. Lett.}\ }\textbf
  {\bibinfo {volume} {95}},\ \bibinfo {pages} {137204} (\bibinfo {year}
  {2005})}\BibitemShut {NoStop}%
\bibitem [{\citenamefont {Thonhauser}\ \emph {et~al.}(2005)\citenamefont
  {Thonhauser}, \citenamefont {Ceresoli}, \citenamefont {Vanderbilt},\ and\
  \citenamefont {Resta}}]{Thonhauser2005}%
  \BibitemOpen
  \bibfield  {author} {\bibinfo {author} {\bibfnamefont {T.}~\bibnamefont
  {Thonhauser}}, \bibinfo {author} {\bibfnamefont {D.}~\bibnamefont
  {Ceresoli}}, \bibinfo {author} {\bibfnamefont {D.}~\bibnamefont
  {Vanderbilt}},\ and\ \bibinfo {author} {\bibfnamefont {R.}~\bibnamefont
  {Resta}},\ }\bibfield  {title} {\bibinfo {title} {Orbital magnetization in
  periodic insulators},\ }\href {https://doi.org/10.1103/PhysRevLett.95.137205}
  {\bibfield  {journal} {\bibinfo  {journal} {Phys. Rev. Lett.}\ }\textbf
  {\bibinfo {volume} {95}},\ \bibinfo {pages} {137205} (\bibinfo {year}
  {2005})}\BibitemShut {NoStop}%
\bibitem [{\citenamefont {Xiao}\ \emph {et~al.}(2010)\citenamefont {Xiao},
  \citenamefont {Chang},\ and\ \citenamefont {Niu}}]{XiaoChangNiu2010}%
  \BibitemOpen
  \bibfield  {author} {\bibinfo {author} {\bibfnamefont {D.}~\bibnamefont
  {Xiao}}, \bibinfo {author} {\bibfnamefont {M.-C.}\ \bibnamefont {Chang}},\
  and\ \bibinfo {author} {\bibfnamefont {Q.}~\bibnamefont {Niu}},\ }\bibfield
  {title} {\bibinfo {title} {Berry phase effects on electronic properties},\
  }\href {https://doi.org/10.1103/RevModPhys.82.1959} {\bibfield  {journal}
  {\bibinfo  {journal} {Rev. Mod. Phys.}\ }\textbf {\bibinfo {volume} {82}},\
  \bibinfo {pages} {1959} (\bibinfo {year} {2010})}\BibitemShut {NoStop}%
\bibitem [{\citenamefont {Gao}\ \emph {et~al.}(2015)\citenamefont {Gao},
  \citenamefont {Yang},\ and\ \citenamefont {Niu}}]{Gao2015}%
  \BibitemOpen
  \bibfield  {author} {\bibinfo {author} {\bibfnamefont {Y.}~\bibnamefont
  {Gao}}, \bibinfo {author} {\bibfnamefont {S.~A.}\ \bibnamefont {Yang}},\ and\
  \bibinfo {author} {\bibfnamefont {Q.}~\bibnamefont {Niu}},\ }\bibfield
  {title} {\bibinfo {title} {Geometrical effects in orbital magnetic
  susceptibility},\ }\href {https://doi.org/10.1103/PhysRevB.91.214405}
  {\bibfield  {journal} {\bibinfo  {journal} {Phys. Rev. B}\ }\textbf {\bibinfo
  {volume} {91}},\ \bibinfo {pages} {214405} (\bibinfo {year}
  {2015})}\BibitemShut {NoStop}%
\bibitem [{\citenamefont {Kang}\ \emph {et~al.}(2025)\citenamefont {Kang},
  \citenamefont {Wang},\ and\ \citenamefont {Vafek}}]{Kang2025}%
  \BibitemOpen
  \bibfield  {author} {\bibinfo {author} {\bibfnamefont {J.}~\bibnamefont
  {Kang}}, \bibinfo {author} {\bibfnamefont {M.}~\bibnamefont {Wang}},\ and\
  \bibinfo {author} {\bibfnamefont {O.}~\bibnamefont {Vafek}},\ }\href@noop {}
  {\bibinfo {title} {Orbital magnetization and magnetic susceptibility of
  interacting electrons}} (\bibinfo {year} {2025}),\ \Eprint
  {https://arxiv.org/abs/2509.20626} {arXiv:2509.20626 [cond-mat.str-el]}
  \BibitemShut {NoStop}%
\bibitem [{\citenamefont {Chen}\ and\ \citenamefont
  {Song}(2026)}]{ChenSong2026}%
  \BibitemOpen
  \bibfield  {author} {\bibinfo {author} {\bibfnamefont {X.}~\bibnamefont
  {Chen}}\ and\ \bibinfo {author} {\bibfnamefont {Z.-D.}\ \bibnamefont
  {Song}},\ }\href@noop {} {\bibinfo {title} {Orbital magnetization of
  interacting electrons}} (\bibinfo {year} {2026}),\ \Eprint
  {https://arxiv.org/abs/2602.02478} {arXiv:2602.02478 [cond-mat.str-el]}
  \BibitemShut {NoStop}%
\bibitem [{\citenamefont {Ye}(2026)}]{Ye2026}%
  \BibitemOpen
  \bibfield  {author} {\bibinfo {author} {\bibfnamefont {M.}~\bibnamefont
  {Ye}},\ }\href@noop {} {\bibinfo {title} {A quantum many-body approach for
  orbital magnetism in correlated multiband electron systems}} (\bibinfo {year}
  {2026}),\ \Eprint {https://arxiv.org/abs/2601.14372} {arXiv:2601.14372
  [cond-mat.str-el]} \BibitemShut {NoStop}%
\bibitem [{\citenamefont {Huang}(2026)}]{Huang2026}%
  \BibitemOpen
  \bibfield  {author} {\bibinfo {author} {\bibfnamefont {C.}~\bibnamefont
  {Huang}},\ }\href@noop {} {\bibinfo {title} {Orbital magnetization from
  uniform and periodic magnetic fields}} (\bibinfo {year} {2026}),\ \Eprint
  {https://arxiv.org/abs/2605.26889} {arXiv:2605.26889} \BibitemShut {NoStop}%
\bibitem [{\citenamefont {Liu}\ \emph {et~al.}(2026)\citenamefont {Liu},
  \citenamefont {Wang}, \citenamefont {Chen}, \citenamefont {Zhang},
  \citenamefont {Cao},\ and\ \citenamefont {Xiao}}]{Liu2026}%
  \BibitemOpen
  \bibfield  {author} {\bibinfo {author} {\bibfnamefont {X.}~\bibnamefont
  {Liu}}, \bibinfo {author} {\bibfnamefont {C.}~\bibnamefont {Wang}}, \bibinfo
  {author} {\bibfnamefont {H.}~\bibnamefont {Chen}}, \bibinfo {author}
  {\bibfnamefont {X.-W.}\ \bibnamefont {Zhang}}, \bibinfo {author}
  {\bibfnamefont {T.}~\bibnamefont {Cao}},\ and\ \bibinfo {author}
  {\bibfnamefont {D.}~\bibnamefont {Xiao}},\ }\bibfield  {title} {\bibinfo
  {title} {Orbital magnetization of correlated states in twisted bilayer
  transition metal dichalcogenides},\ }\href
  {https://doi.org/10.1103/k46f-8m8t} {\bibfield  {journal} {\bibinfo
  {journal} {Phys. Rev. Lett.}\ }\textbf {\bibinfo {volume} {136}},\ \bibinfo
  {pages} {166606} (\bibinfo {year} {2026})}\BibitemShut {NoStop}%
\bibitem [{\citenamefont {Groenewold}(1946)}]{Groenewold1946}%
  \BibitemOpen
  \bibfield  {author} {\bibinfo {author} {\bibfnamefont {H.~J.}\ \bibnamefont
  {Groenewold}},\ }\bibfield  {title} {\bibinfo {title} {On the principles of
  elementary quantum mechanics},\ }\href
  {https://doi.org/10.1016/S0031-8914(46)80059-4} {\bibfield  {journal}
  {\bibinfo  {journal} {Physica}\ }\textbf {\bibinfo {volume} {12}},\ \bibinfo
  {pages} {405} (\bibinfo {year} {1946})}\BibitemShut {NoStop}%
\bibitem [{\citenamefont {Moyal}(1949)}]{Moyal1949}%
  \BibitemOpen
  \bibfield  {author} {\bibinfo {author} {\bibfnamefont {J.~E.}\ \bibnamefont
  {Moyal}},\ }\bibfield  {title} {\bibinfo {title} {Quantum mechanics as a
  statistical theory},\ }\href {https://doi.org/10.1017/S0305004100000487}
  {\bibfield  {journal} {\bibinfo  {journal} {Proc. Cambridge Philos. Soc.}\
  }\textbf {\bibinfo {volume} {45}},\ \bibinfo {pages} {99} (\bibinfo {year}
  {1949})}\BibitemShut {NoStop}%
\bibitem [{\citenamefont {St\v{r}eda}(1982)}]{Streda1982}%
  \BibitemOpen
  \bibfield  {author} {\bibinfo {author} {\bibfnamefont {P.}~\bibnamefont
  {St\v{r}eda}},\ }\bibfield  {title} {\bibinfo {title} {Theory of quantised
  hall conductivity in two dimensions},\ }\href
  {https://doi.org/10.1088/0022-3719/15/22/005} {\bibfield  {journal} {\bibinfo
   {journal} {J. Phys. C: Solid State Phys.}\ }\textbf {\bibinfo {volume}
  {15}},\ \bibinfo {pages} {L717} (\bibinfo {year} {1982})}\BibitemShut
  {NoStop}%
\bibitem [{\citenamefont {Zhu}\ and\ \citenamefont {Huang}(2026)}]{Zhu2026}%
  \BibitemOpen
  \bibfield  {author} {\bibinfo {author} {\bibfnamefont {J.}~\bibnamefont
  {Zhu}}\ and\ \bibinfo {author} {\bibfnamefont {C.}~\bibnamefont {Huang}},\
  }\bibfield  {title} {\bibinfo {title} {Orbital magnetization and streda
  formula of interacting electrons in the mean-field approximation},\ }\href
  {https://doi.org/10.1103/w49b-25pd} {\bibfield  {journal} {\bibinfo
  {journal} {Phys. Rev. B}\ }\textbf {\bibinfo {volume} {113}},\ \bibinfo
  {pages} {205114} (\bibinfo {year} {2026})}\BibitemShut {NoStop}%
\bibitem [{\citenamefont {Lian}\ \emph {et~al.}(2021)\citenamefont {Lian},
  \citenamefont {Xie},\ and\ \citenamefont {Bernevig}}]{Lian2021}%
  \BibitemOpen
  \bibfield  {author} {\bibinfo {author} {\bibfnamefont {B.}~\bibnamefont
  {Lian}}, \bibinfo {author} {\bibfnamefont {F.}~\bibnamefont {Xie}},\ and\
  \bibinfo {author} {\bibfnamefont {B.~A.}\ \bibnamefont {Bernevig}},\
  }\bibfield  {title} {\bibinfo {title} {{Open momentum space method for the
  Hofstadter butterfly and the quantized Lorentz susceptibility}},\ }\href
  {https://doi.org/10.1103/physrevb.103.l161405} {\bibfield  {journal}
  {\bibinfo  {journal} {Phys. Rev. B}\ }\textbf {\bibinfo {volume} {103}},\
  \bibinfo {pages} {l161405} (\bibinfo {year} {2021})}\BibitemShut {NoStop}%
\bibitem [{\citenamefont {Raoux}\ \emph {et~al.}(2015)\citenamefont {Raoux},
  \citenamefont {Pi\'echon}, \citenamefont {Fuchs},\ and\ \citenamefont
  {Montambaux}}]{Raoux2015}%
  \BibitemOpen
  \bibfield  {author} {\bibinfo {author} {\bibfnamefont {A.}~\bibnamefont
  {Raoux}}, \bibinfo {author} {\bibfnamefont {F.}~\bibnamefont {Pi\'echon}},
  \bibinfo {author} {\bibfnamefont {J.-N.}\ \bibnamefont {Fuchs}},\ and\
  \bibinfo {author} {\bibfnamefont {G.}~\bibnamefont {Montambaux}},\ }\bibfield
   {title} {\bibinfo {title} {Orbital magnetism in coupled-bands models},\
  }\href {https://doi.org/10.1103/PhysRevB.91.085120} {\bibfield  {journal}
  {\bibinfo  {journal} {Phys. Rev. B}\ }\textbf {\bibinfo {volume} {91}},\
  \bibinfo {pages} {085120} (\bibinfo {year} {2015})}\BibitemShut {NoStop}%
\bibitem [{\citenamefont {Wu}\ \emph {et~al.}(2019)\citenamefont {Wu},
  \citenamefont {Lovorn}, \citenamefont {Tutuc}, \citenamefont {Martin},\ and\
  \citenamefont {MacDonald}}]{Wu2019}%
  \BibitemOpen
  \bibfield  {author} {\bibinfo {author} {\bibfnamefont {F.}~\bibnamefont
  {Wu}}, \bibinfo {author} {\bibfnamefont {T.}~\bibnamefont {Lovorn}}, \bibinfo
  {author} {\bibfnamefont {E.}~\bibnamefont {Tutuc}}, \bibinfo {author}
  {\bibfnamefont {I.}~\bibnamefont {Martin}},\ and\ \bibinfo {author}
  {\bibfnamefont {A.~H.}\ \bibnamefont {MacDonald}},\ }\bibfield  {title}
  {\bibinfo {title} {Topological insulators in twisted transition metal
  dichalcogenide homobilayers},\ }\href
  {https://doi.org/10.1103/PhysRevLett.122.086402} {\bibfield  {journal}
  {\bibinfo  {journal} {Phys. Rev. Lett.}\ }\textbf {\bibinfo {volume} {122}},\
  \bibinfo {pages} {086402} (\bibinfo {year} {2019})}\BibitemShut {NoStop}%
\bibitem [{\citenamefont {Wang}\ and\ \citenamefont {Vafek}(2024)}]{Wang2024}%
  \BibitemOpen
  \bibfield  {author} {\bibinfo {author} {\bibfnamefont {X.}~\bibnamefont
  {Wang}}\ and\ \bibinfo {author} {\bibfnamefont {O.}~\bibnamefont {Vafek}},\
  }\bibfield  {title} {\bibinfo {title} {{Theory of Correlated Chern Insulators
  in Twisted Bilayer Graphene}},\ }\href
  {https://doi.org/10.1103/physrevx.14.021042} {\bibfield  {journal} {\bibinfo
  {journal} {Phys. Rev. X}\ }\textbf {\bibinfo {volume} {14}},\ \bibinfo
  {pages} {021042} (\bibinfo {year} {2024})}\BibitemShut {NoStop}%
\end{thebibliography}%

\clearpage
\onecolumngrid
\vspace{1.5\baselineskip}
\begin{center}
	\textbf{\large End Matter}
\end{center}
\vspace{0.5\baselineskip}
\twocolumngrid

\appendix

\appxsection{Hartree-Fock self-consistent equation}{App:selfcons}
In this appendix, we derive the HF self-consistent equation.
In the Hartree-Fock approximation, the interaction becomes Hartree and Fock terms, yielding the following mean-field Hamiltonian
\begin{align}
	&\hat H^{\text{MF}}=\hat h+\int\dd[2]{r}\dd[2]{r'}V(\vb r-\vb r')\expval{n(\vb r')}n(\vb r)\\
	&-\sum_{\alpha,\beta}\int\dd[2]{r}\dd[2]{r'}V(\vb r-\vb r')\expval{\psi^\dag_{\alpha}(\vb r')\psi_{\beta}(\vb r)}\psi^\dag_{\beta}(\vb r)\psi_{\alpha}(\vb r'), \nonumber
\end{align}
where the second and third terms are respectively the Hartree and Fock terms, and $n(\vb r)=\sum_\alpha \psi^\dag_{\alpha}(\vb r)\psi_{\alpha}(\vb r)$ is the total density.
If the ground state has continuous or discrete translation symmetry, the Hartree term becomes
\begin{align}
	\text{Hartree\ term}=\mathcal{A}^{-1}\sum_{\vb g\in\Lambda} \tilde V(g) e^{i\vb g\vdot \hat{\vb r}}\expval{e^{-i\vb g\vdot \hat{\vb r}}},
\end{align}
where $\Lambda=\{\vb 0\}$ for translation-invariant systems, and $\Lambda$ is the set of all reciprocal vectors for periodic systems. As the $\vb g=\vb 0$ contribution is always canceled by the positive charge background, we set $\tilde V(q=0)=0$ henceforth. 

The Fock term can be simplified as
\begin{align}\label{Eq:Fock1}
	\text{Fock\ term}&=\sum_{\alpha,\beta}\int\dd[2]q\tilde V(q)\int\dd[2]{r}\dd[2]{r'}e^{i\vb q\vdot(\vb r-\vb r')}\nonumber\\
	&\quad\quad\quad\quad\times\expval{\psi^\dag_{\alpha}(\vb r')\psi_{\beta}(\vb r)}\psi^\dag_{\beta}(\vb r)\psi_{\alpha}(\vb r')\nonumber\\
	&=\int\dd[2]q\tilde V(q)e^{i\vb q\vdot \hat{\vb r}}\hat N e^{-i\vb q\vdot \hat{\vb r}},
\end{align}
because in the first-quantization language, we have
\begin{align}
	&\sum_{\alpha,\beta}\int\dd[2]{r}\dd[2]{r'}\expval{\psi^\dag_{\alpha}(\vb r)\psi_{\beta}(\vb r')}\psi^\dag_{\beta}(\vb r')\psi_{\alpha}(\vb r)\nonumber\\
	&=\sum_{\alpha,\beta}\int\dd[2]{r}\dd[2]{r'}\mel{\vb r\alpha}{\hat N}{\vb r'\beta}\dyad{\vb r\alpha}{\vb r'\beta}=\hat N,
\end{align}
where $\ket{\vb r\alpha}$ is the eigenstate of position operator of flavor $\alpha$. Substituting it into Eq.~\eqref{Eq:Fock1}, we obtain the result in the main text.

\appxsection{Reverse Peierls substitution}{App:peierls}
In this appendix, we prove the existence and uniqueness of reverse Peierls substitution. Because the canonical momenta $\hat{\vb*\pi}$ commute with the guiding center momenta $\hat{\vb Q}$, the Hilbert space can be decomposed into the tensor product of two Landau level systems. In the symmetric gauge, the canonical momenta generate the physical Landau levels, while the guiding center momenta generate the angular momentum.
The Peierls substitution is a Weyl transform, whose unique inverse is the Wigner map. Thus, for any operator $\hat A$, there exists a function $f(\vb s,\vb s')$ satisfying
\begin{align}
	\hat A &=\int\dd[2]{s}\dd[2]{s'} f(\vb s,\vb s')e^{i\vb s\vdot \hat{\vb*\pi}+i\vb s'\vdot \hat{\vb Q}}\nonumber\\
	&=\int\dd[2]{s}\dd[2]{s'} \tilde f(\vb s,\vb s')e^{i\vb s\vdot \hat{\vb*\pi}}e^{i\vb s'\vdot \hat{\vb r}}
\end{align}
with $\tilde f(\vb s,\vb s')=B^2 f(\vb s-\vb s'\cp\vb z/B,\vb s'\cp\vb z/B)e^{i\vb s\vdot\vb s'}$.

If $\hat A$ is translation-invariant, $\hat A$ is only a function of $\pi$ and $f(\vb s,\vb s')$ only takes value at $\vb s'=\vb 0$.
If $\hat A$ is periodic, then $\tilde f(\vb s,\vb s')=\tilde f(\vb s,\vb s')e^{i\vb s'\vdot \vb R}$ for all lattice vectors $\vb R$ in real space. Therefore, $\tilde f(\vb s,\vb s')$ only takes value at $\vb s'=\vb g\in\Lambda$, and correspondingly, we have
\begin{align}
	\hat A&=\sum_{\vb g\in \Lambda}\int\dd[2]s f(\vb s,\vb g)e^{i\vb {s}\vdot \hat{\vb*\pi}}e^{i\vb g\vdot\hat{\vb r}}
	=\sum_{\vb g\in\Lambda}A_{\vb g}(\hat{\vb*\pi})e^{i\vb g\vdot\hat{\vb r}},
\end{align}
where $A_{\vb g}(\hat{\vb*\pi})=\int\dd[2]s f(\vb s,\vb g)e^{i\vb {s}\vdot \hat{\vb*\pi}}$.

\appxsection{Density change at finite magnetic fields}{App:density}
In this appendix, we will use Eq.~\eqref{Eq:algebra} to derive the analytic expressions of $\delta N_{\pi}$ and $\delta N_{\delta X}$, which are
\begin{align}\label{Eq:appA}
 &\delta N_{\pi}(\vb k)=-\frac{i}{2}\sum_{\alpha,\beta}\dyad{\alpha}{\beta}\sum_{\gamma}K_{f_{\text{FD}}}(\varepsilon_\alpha,\varepsilon_\gamma,\varepsilon_\beta)\vb z\vdot(\vb v_{\alpha\gamma}\cp \vb v_{\gamma\beta}),\nonumber\\
 &\delta N_{\delta X}(\vb k)=\sum_{\varepsilon_\alpha\neq\varepsilon_\beta}\dyad{\alpha}{\beta}\frac{f_\alpha-f_\beta}{\varepsilon_\alpha-\varepsilon_\beta}\mel{\alpha}{\delta X}{\beta}\nonumber\\
 &\qquad\qquad\quad+\sum_{\varepsilon_\alpha=\varepsilon_\beta} \dyad{\alpha}{\beta} f'_\alpha\mel{\alpha}{\delta X}{\beta},
\end{align}
where $\ket{\alpha}$ is the eigenstate of $H^{\text{MF}}_0(\vb k)$ with eigenvalue $\varepsilon_\alpha$, $\vb v_{\alpha\beta}=\mel{\alpha}{\partial H^{\text{MF}}_0(\vb k)}{\beta}$, $f_\alpha=f_{\text{FD}}(\varepsilon_\alpha)$, and for a univariate function $f$ and $a_1\neq a_2\neq a_3\neq a_1$, we define
\begin{align}
	K_f(a_1,a_2,a_3)&=\frac12\sum_{\sigma\in S_3}\frac{f(a_{\sigma(1)})}{(a_{\sigma(2)}-a_{\sigma(1)})(a_{\sigma(3)}-a_{\sigma(1)})}\nonumber\\
	K_f(a_2,a_1,a_1)&=K_f(a_1,a_2,a_1)=K_f(a_1,a_1,a_2)\nonumber\\
	&=\frac{f(a_2)-f(a_1)}{(a_2-a_1)^2}-\frac{f'(a_1)}{a_2-a_1}\nonumber\\
	K_f(a_1,a_1,a_1)&=f''(a_1)/2, 
\end{align}
with $S_3$ denoting the permutation group of 3.
Following Eq.~\eqref{Eq:pertgen}, both sums in $\delta N_{\delta X}$ are keyed on eigenvalues rather than state labels, so that degenerate pairs $\alpha\neq\beta$ enter the second sum and $\delta N_{\delta X}$ is basis independent within a multiplet.

One can inductively prove for all integers $a\geq 2$ that
\begin{align}\label{Eq:power}
	&\lim_{B\to 0}\frac{\mel{\alpha}{\mathcal{R}[H(\hat{\vb*\pi})^{a}]-H^{a}}{\beta}}{B}\nonumber\\
	&=\lim_{B\to 0}\frac{\mel{\alpha}{(\mathcal{R}[H(\hat{\vb*\pi})^{a-1}]-H^{a-1})H}{\beta}}{B}-\frac i2\vb z\vdot (\partial H^{a-1}\cp \vb v)\nonumber\\
	&=-\frac i2\sum_{\gamma} \vb z\vdot(\vb v_{\alpha\gamma}\cp \vb v_{\gamma\beta}) \sum_{m=0}^{a-2}\sum_{n=0}^{a-m-2}\varepsilon_\alpha^{m}  \varepsilon_\gamma^n  \varepsilon_\beta^{a-m-n-2}\nonumber\\
	&=-\frac i2\sum_{\gamma} \vb z\vdot(\vb v_{\alpha\gamma}\cp \vb v_{\gamma\beta}) K_{x^a}(\varepsilon_\alpha,\varepsilon_\gamma,\varepsilon_\beta),
\end{align}
where all $\vb k$ arguments and the superscript and subscript of $H^{\text{MF}}_0$ are omitted for convenience, $\vb v=\partial H$, and the last equality is obtained by calculating the sum of the geometric sequence. Moreover, as $K_{x^0}(a_1,a_2,a_3)=K_{x^1}(a_1,a_2,a_3)=0$, Eq.~\eqref{Eq:power} also holds for $a=0,1$.
Hence, Eq.~\eqref{Eq:power} also holds if the power function $x^a$ is replaced with any polynomial $P(x)$. 
Additionally, because functions can generally be approximated by polynomials, Eq.~\eqref{Eq:power} also holds if the power function $x^a$ is replaced with the Fermi-Dirac distribution, which gives the $\delta N_{\pi}$ in Eq.~\eqref{Eq:appA}.

Similarly, it is easy to prove that for polynomials $P(x)$
\begin{align}\label{Eq:pertgen}
	&\lim_{B\to 0}\frac{\mel{\alpha}{P(H+B\delta X)-P(H)}{\beta}}B\nonumber\\
	&=
	\begin{cases}
    \frac{P(\varepsilon_{\alpha})-P(\varepsilon_{\beta})}{\varepsilon_{\alpha}-\varepsilon_{\beta}}\mel{\alpha}{\delta X}{\beta} & \text{if } \varepsilon_{\alpha}\neq\varepsilon_{\beta}, \\
    P'(\varepsilon_{\alpha})\mel{\alpha}{\delta X}{\beta}   & \text{if } \varepsilon_{\alpha}=\varepsilon_{\beta}.
\end{cases}
\end{align}
By the same argument, we can replace the polynomial $P(x)$ with the Fermi-Dirac distribution $f_{\text{FD}}(x)$.

\appxsection{Algebra in the extended Hilbert space}{App:algebra}
In this appendix, we prove that the Hilbert-space extension introduced in the main text preserves the algebra of periodic operators. 
We only need to prove that 
$\mathcal M[F(\hat{\vb*\pi})e^{i\vb g_1\vdot\hat{\vb r}}]\mathcal M[G(\hat{\vb*\pi})e^{i\vb g_2\vdot\hat{\vb r}}]
	=\mathcal M[F(\hat{\vb*\pi})G(\hat{\vb*\pi}-\vb g_1)e^{i\vb{(g_1+g_2)}\vdot\hat{\vb r}}]$. 
This can be proven by noting that
\begin{align}
	&\quad \mathcal M[F(\hat{\vb*\pi})G(\hat{\vb*\pi}-\vb g_1)e^{i\vb{(g_1+g_2)}\vdot\hat{\vb r}}]\nonumber\\
	&=\sum_{\vb g'\in\Lambda}F(\hat{\vb*\pi}+\vb g')G(\hat{\vb*\pi}+\vb g'-\vb g_1 )\dyad{\vb g'}{\vb g'-\vb{g_1-g_2}}\nonumber\\
	&=\sum_{\vb g',\vb g''\in\Lambda}F(\hat{\vb*\pi}+\vb g')\dyad{\vb g'}{\vb g'-\vb{g}_1}G(\hat{\vb*\pi}+\vb g'' )\dyad{\vb g''}{\vb g''-\vb{g_2}}\nonumber\\
	&=\mathcal M[F(\hat{\vb*\pi})e^{i\vb g_1\vdot\hat{\vb r}}]\mathcal M[G(\hat{\vb*\pi})e^{i\vb g_2\vdot\hat{\vb r}}].
\end{align}

\appxsection{Calculation of $\tilde \chi_{\delta X}$}{App:chidX} In this appendix, we derive the formula for $\tilde \chi_{\delta X}$. The mean field magnetization is given by
\begin{align}
	M^{\text{HF}}=\sum_\alpha\int\frac{\dd[2]{k}}{(2\pi)^2}&[f_\alpha m_\alpha +g_\alpha \Omega_\alpha ], 
\end{align}
where $g_\alpha=T\ln[1+e^{(\mu-\varepsilon_\alpha)/T}]$. 
According to the first-order perturbation theory, the first-order change of energy and eigenstate is $\delta \varepsilon_\alpha=\delta X_\alpha$ and $\ket{\delta \alpha}=\sum_{\beta\neq\alpha}\frac{\ket{\beta}\mel{\beta}{\delta X}{\alpha}}{\varepsilon_\alpha-\varepsilon_\beta}$.
Thus, we have
\begin{widetext}
\begin{align}
	\tilde \chi_{\delta X}&=\sum_\alpha\int\frac{\dd[2]{k}}{(2\pi)^2}[f_\alpha \delta m_\alpha +g_\alpha \delta \Omega_\alpha + \delta X_\alpha(f'_\alpha m_\alpha -f_\alpha \Omega_\alpha)]\nonumber,\\
	\delta m_\alpha&=
	\text{Im}[\mel{\delta_x\alpha}{(H_0^{\text{MF}}-\varepsilon_\alpha)}{\partial_y \alpha}
	+\mel{\partial_x \alpha}{(H_0^{\text{MF}}-\varepsilon_\alpha)}{\delta_y\alpha}
	+\mel{\partial_x \alpha}{(\delta X-\delta X_\alpha)}{\partial_y \alpha}]\nonumber,\\
	\delta \Omega_\alpha&=-2\text{Im}[\braket{\delta_x\alpha}{\partial_y \alpha}+\braket{\partial_x \alpha}{\delta_y\alpha}], 
\end{align}
where $\ket{\partial_{x/y} \alpha}=\sum_{\beta\neq\alpha}\frac{\ket{\beta}\mel{\beta}{v_{x/y}}{\alpha}}{\varepsilon_\alpha-\varepsilon_\beta}$, and $\ket{\delta_{x/y} \alpha}$ denotes the first-order change of $\ket{\partial_{x/y} \alpha}$, which is given by
\begin{align}
	\ket{\delta_{x/y} \alpha}=\sum_{\beta\neq\alpha}\qty[\ket{\delta\beta}\frac{\mel{\beta}{v_{x/y}}{\alpha}}{\varepsilon_\alpha-\varepsilon_\beta}+\ket{\beta}\frac{\mel{\delta\beta}{v_{x/y}}{\alpha}+\mel{\beta}{\partial_{x/y}(\delta X)}{\alpha}+\mel{\beta}{v_{x/y}}{\delta\alpha}}{\varepsilon_\alpha-\varepsilon_\beta}
	-\ket{\beta}\frac{\mel{\beta}{v_{x/y}}{\alpha}(\delta X_\alpha-\delta X_\beta)}{(\varepsilon_\alpha-\varepsilon_\beta)^2}]. \notag
\end{align}
\end{widetext}

\appxsection{Two toy models}{App:models}
For the translation-invariant model in Eq.~\eqref{Eq:TI model}, we take $(v,m,\Delta,\lambda,U)=(1,1,0.2,1,2\pi^2)$. The chemical potential is fixed at $\mu=-1.2$ for calculations at both zero and finite fields. Moreover, we introduce a very low temperature, $T=0.01$, to reduce quantum oscillations. Particularly, the grand potential density at finite temperature is given by
\begin{align}
	\Omega&=\frac1A\{\trace[\hat N(\hat h-\mu)]+\trace[\hat NV_{\text{MF}}(\hat N)]/2\nonumber\\
	&\quad\quad+T\trace[\hat N\ln \hat N+(1-\hat N){\ln} (1-\hat N)]\}.
\end{align}

For the continuum model of tMoTe$_2$  in Eq.~\eqref{Eq:moire model}, its \moire potential is given by
\begin{align}
	V_l(\vb r)&=-2V\sum_{j=1}^3 \cos(\vb G_j\vdot \vb r+l\psi),\\
	T(\vb r)&=-w(1+e^{-i\vb G_2\vdot\vb r}+e^{i\vb G_3\vdot\vb r}),
\end{align}
where $l=1$ ($l=-1$) for the bottom (top) layer, $\vb G_1=k_\theta[1,0],\vb G_2=k_\theta[-1/2,\sqrt3/2]$ are the basis vectors of reciprocal vectors of tMoTe$_2$, $\vb G_3=-\vb G_1-\vb G_2$. The $K$ valleys of the two layers are $\vb*\kappa_b=-(2\vb G_1+\vb G_2)/3$ and $\vb*\kappa_t=(\vb G_2-\vb G_1)/3$. Here, $k_\theta=8\pi\sin(\theta/2)/(\sqrt3a_0)$ describes the length scale of the system, where $a_0$ is the lattice constant of MoTe$_2$, and $\theta$ is the twist angle. Adopting the parameters in Ref.~\cite{Wu2019}, we take $(a_0,m,V,\psi,w)=(\SI{3.472}{\angstrom},0.62m_e,\SI{8}{meV},-89.6^\circ,\SI{-8.5}{meV})$, where $m_e$ is the bare electron mass, and $\theta=1.7^\circ$.

\end{document}